\documentclass[11pt,a4paper]{article}
\usepackage{amssymb, amsmath,framed}
\usepackage[colorlinks]{hyperref}
\usepackage[margin=1in]{geometry}
\newcommand{\rem}[1]{}
\newcommand{\de}{{\rm d}}

\newcommand{\bq}{{\mathbf{x}}}
\newcommand{\bz}{{\mathbf{z}}}
\newcommand{\br}{{\mathbf{r}}}
\newcommand{\bc}{{\mathbf{c}}}

\newcommand{\bM}{{\mathbf{M}}}
\newcommand{\bK}{{\mathbf{K}}}
\newcommand{\bE}{{\mathbf{E}}}
\newcommand{\bA}{{\mathbf{A}}}
\newcommand{\bB}{{\mathbf{B}}}

\newcommand{\ba}{{\boldsymbol{a}}}

\newcommand{\bcalX}{{\mathbf{X}}}
\newcommand{\bu}{{\mathbf{u}}}
\newcommand{\bv}{{\mathbf{v}}}
\newcommand{\bU}{{\boldsymbol{U}}}
\newcommand{\bb}{{\boldsymbol{b}}}
\newcommand{\bgamma}{{\boldsymbol{\gamma}}}

\newcommand{\bW}{{\boldsymbol{W}}}

\newcommand{\bX}{{\mathbf{X}}}

\newcommand{\bw}{{\mathbf{w}}}

\newcommand{\pa}{\partial}
\newcommand{\vp}{{v_\parallel}}
\newcommand{\bJ}{{\mathbf{J}}}

\newcommand{\comment}[1]{\vspace{1 mm}\par
\marginpar{\large\underline{}}\noindent
\framebox{\begin{minipage}[c]{0.95 \textwidth}
\rm\tt #1 \end{minipage}}\vspace{2 mm}\par}

\begin{document}

\title{Variational mean-fluctuation splitting and drift-fluid models}
\author{Cesare Tronci \smallskip 
\\ 
\small
\it Department of Mathematics, University of Surrey, Guildford, United Kingdom
\\
\small
\it %Numerical Methods in Plasma Physics Division, 
Numerical Methods Division, Max Planck Institute for Plasma Physics, Garching, Germany}
\date{$\,$}

\maketitle

\begin{abstract}
After summarizing the variational approach to  splitting mean flow and fluctuation kinetics in the standard Vlasov theory, the same method is applied to the drift-kinetic equation from Littlejohn's theory of guiding-center motion. This process sheds a new light on drift-ordered fluid (drift-fluid) models, whose anisotropic pressure tensor is then considered in detail. In addition, current drift-fluid models are completed by the insertion of magnetization  terms ensuring momentum conservation. Magnetization currents are also shown to lead to challenging aspects when drift-fluid models are coupled to Maxwell's equations for the evolution of the electromagnetic field. In order to overcome these difficulties, a simplified guiding-center theory is proposed along with its possible applications to hybrid kinetic-fluid models. 

\end{abstract}

{\small
\tableofcontents
}

\section{Introduction}

\subsection{The mean-fluctuation splitting in Vlasov kinetic theory}

Within the Vlasov kinetic theory of charged particles, fluid models are usually obtained by well-established moment methods. In this process, it is convenient to split the dynamics of the mean flow velocity from the kinetics of velocity fluctuations. More specifically, one introduces  the mean-flow velocity $\bU(\bq,t)$ defined by $\bU(\bq,t)\int\! F(\bq,\bv,t)\,\de^3v=\int\! \bv F(\bq,\bv,t)\,\de^3v$ and rewrites the Vlasov kinetic equation  in terms of the following system
\begin{align}\label{EQ1}
&m n\left(\frac{\partial\bU}{\partial t}+\bU\cdot\nabla\bU\right)=-\operatorname{div}\Bbb{P}+qn\bE+qn\bU\times\bB
\\\label{EQ2}
&\frac{\partial \widetilde{F}}{\partial t}+\left(\bc+\bU\right)\cdot\frac{\partial \widetilde{F}}{\partial \bq}+\left(\frac{q}{m}\bE+\frac{q}{m}(\bc+\bU)\times\bB-(\bc+\bU)\cdot\nabla\bU-\frac{\partial \bU}{\partial t}\right)\cdot\frac{\partial \widetilde{F}}{\partial \bc}=0\,,
\end{align}
where 
\begin{equation}
n=\int\! \widetilde{F}(\bq,\bc,t)\,\de^3c
\,,\qquad
\Bbb{P} = m \int\! \bc\bc\widetilde{F}(\bq,\bc,t)\,\de^3c
\label{moments}
\end{equation}
are the zero-th and second kinetic moment, respectively, $\bc=\bv-\bU$ is the fluctuation velocity coordinate, and $\widetilde{F}(\bq,\bc,t)={F}(\bq,\bv,t)$ is the probability density in the Eulerian frame moving with the mean-flow velocity.
This system \eqref{EQ1}-\eqref{EQ2} is the most fundamental example of a hybrid kinetic-fluid model, a class of models widely studied in the computational physics of magnetized plasmas \cite{PaBeFuTaStSu}. A fluid model is obtained from \eqref{EQ1}-\eqref{EQ2} by making assumptions on the pressure tensor (density) $\Bbb{P}$. Most often, these assumptions make use of a {\it closure}, that is an ansatz on the form of the Vlasov distribution depending on the first three moments. The most common class of models is obtained by assuming a  zero-mean Gaussian distribution in the relative velocity $\bc$, with (usually diagonal) variance matrix given by $\Bbb{P}(\bq,t)$ and zero-th velocity moment by $n(\bq,t)$.

The variational structure of the mean-fluctuation splitting \eqref{EQ1}-\eqref{EQ2} was exploited in \cite{Tronci2014} to derive a class of kinetic models for the description of magnetic reconnection in space plasmas. In turn, this variational structure is actually identical to that of hybrid MHD models in  the pressure-coupling scheme \cite{ClBuTr,HoTr2011}.
This paper addresses the question of how the mean-fluctuation splitting applies to Littlejohn's kinetic equation for guiding-center motion \cite{Littlejohn} in order to shed a new light on the formulation of the so-called drift-ordered fluid models, or simply \emph{drift-fluids}. In particular, while drift-fluid models may be obtained from different procedures, here we shall compare our results with certain classes of gyrofluid models \cite{Pfirsch,StSc,StScBr} obtained within a similar variational approach. As we shall see, important features will emerge along the way, such as the role of the magnetization current and the parallel pressure as a source of magnetic reconnection. In addition, when the mean-fluctuation splitting is performed in the presence of the accompanying Maxwell's equations for the electromagnetic field, we shall see that the intricate structure of guiding-center theory leads to drift-ordered models much more involved than in the case of external electromagnetic fields. In order to simplify the treatment, an alternative guiding-center model is proposed along with its possible applications in hybrid kinetic-fluid models.

\subsection{Euler-Poincar\'e  variational structure\label{sec:Vlasov}}

As anticipated above,  the general variational structure of the mean-fluctuation splitting \eqref{EQ1}-\eqref{EQ2} is the same as that underlying certain hybrid MHD models for the description of energetic particle effects \cite{HoTr2011,ClBuTr}. This Euler-Poincar\'e variational structure \cite{CeMaHo,HoMaRa1998,IlLa,Newcomb,SqQiTa} was also exploited in \cite{Tronci2014} to formulate fully kinetic reconnection models requiring lengthscales much smaller than the electron skin depth. These models were motivated by previously established moment truncations \cite{Winske} and were recently discussed in \cite{CaTa}. Similar Euler-Poincar\'e variational structures have also been used recently to approach the kinetic-MHD model \cite{BuSe}.

{\color{black}The Euler-Poincar\'e reduction \cite{HoMaRa1998} of Hamilton's principle for continuum dynamics is a powerful tool relating the celebrated Lagrangian and Eulerian pictures of fluid flows. In this context, while the Lagrangian paths possess arbitrary variations as in standard classical mechanics, the Eulerian quantities obtained upon reduction by the relabeling symmetry possess constrained variations as in Newcomb's early works \cite{Newcomb}. For example, if $\boldsymbol\eta(\boldsymbol{x}_0,t)$ is a Lagrangian fluid path with arbitrary variation $\delta\boldsymbol\eta$, then the Eulerian velocity $\boldsymbol{v}(\boldsymbol{x},t)$ such that $\dot{\boldsymbol\eta}(\boldsymbol{x}_0,t)=\boldsymbol{v}(\boldsymbol{x},t)|_{\boldsymbol{x}=\boldsymbol\eta(\boldsymbol{x}_0,t)}$ has variations $\delta\boldsymbol{v}=\delta(\dot{\boldsymbol\eta}\circ{\boldsymbol\eta}^{-1})$, where $\circ$ denotes composition of functions. A vector calculus exercise shows that $\delta\boldsymbol{v}=\partial_t\boldsymbol\xi+\boldsymbol{v}\cdot\nabla\boldsymbol\xi-\boldsymbol\xi\cdot\nabla\boldsymbol{v}$, where $\boldsymbol{\xi}$ is the arbitrary infinitesimal displacement such that $\delta{\boldsymbol\eta}(\boldsymbol{x}_0,t)=\boldsymbol{\xi}(\boldsymbol{x},t)|_{\boldsymbol{x}=\boldsymbol\eta(\boldsymbol{x}_0,t)}$. In the specific case of the mean-fluctuation splitting \eqref{EQ1}-\eqref{EQ2}, the geometric structure of  its underlying Euler-Poincar\'e variational principle is more involved and it
}
exploits the use of the so-called \emph{tangent lifts} of the fluid path from physical space to phase space, so that Lagrangian fluid paths and particle paths on phase-space can be composed together to realize the change of frame \cite{ThMc} performed in the mean-fluctuation splitting. See \cite{ClBuTr,HoTr2011} for further discussions. Instead of reviewing the geometry underlying the Euler-Poincar\'e variational principle underlying \eqref{EQ1}-\eqref{EQ2}, here we shall simply write the corresponding Euler-Poincar\'e equations as they arise from an arbitrary Lagrangian functional of the type $l(\bU,\bX,F)$. Notice that we have conveniently dropped the tilde on the distribution function $\widetilde{F}(\bq,\bc,t)$ from the previous section. Also,  $\bX(\bq,\bc,t)=\big(\bw(\bq,\bc,t),\ba(\bq,\bc,t)\big)$ is a vector field on phase-space constructed in such a way that, by abusing notation, $\dot\bq=\bw+\bU$ (and also $\dot\bc=\ba+\bc\cdot\nabla\bU$, by the same abuse of notation). Upon denoting $\bz=(\bq,\bc)$, the general form of the Euler-Poincar\'e equations  is derived from the variational principle \cite{HoTr2011}
\[
\delta\int_{t_1}^{t_2} l(\bU,\bX,F)\,\de t=0
\]
upon using the following variations
\begin{align*}
&\delta\bU=\partial_t\boldsymbol\Xi+[\bU,\boldsymbol\Xi]\,,
\\
&\delta{\bX}=\partial_t\mathbf{Y}+[{\bX},{\mathbf{Y}}]+[\bX_{\bU},{\mathbf{Y}}]-[\bX_{\boldsymbol\Xi},{\bX}]\,,
\\
&\delta F=-\nabla_\bz\cdot(F\,{\mathbf{Y}})-\nabla_\bz\cdot(F\,\bX_{\boldsymbol\Xi})
\,.
\end{align*}
Here,  $[\mathbf{P},\mathbf{R}]=(\mathbf{P}\cdot\nabla)\mathbf{R}-(\mathbf{R}\cdot\nabla)\mathbf{P}$ is the vector field commutator, $\boldsymbol\Xi(\bq,t)$ and $\mathbf{Y}(\bq,\bc,t)$ are arbitrary displacement vector fields vanishing at the endpoints $t_{1}$ and $t_2$, while $\bX_\bU=(\bU,\bc\cdot\nabla\bU)$ is the phase-space vector field generated by the tangent lift of the fluid velocity vector field $\bU$ onto phase-space (analogously for $\bX_{\boldsymbol\Xi}$). Eventually, one is left with the following system in terms of functional derivatives:
\begin{align}\label{EP-PCS1}
&\frac{\partial}{\partial t}\frac{\delta l}{\delta
{\bU}}+\pounds_{\bU}\,\frac{\delta l}{\delta \bU}=-
\int\!\left(\!\pounds_{\bcalX}\frac{\delta  l}{\delta
\bcalX}- {F}\nabla_{\bz}\frac{\delta l}{\delta F}\right)_{\!\!\bq}
\!\de^3c
+
\nabla\cdot\!\int\!\bc\left(\!\pounds_{ \bcalX}\frac{\delta
 l}{\delta  \bcalX}-{F}\,\nabla_{\bz}\frac{\delta l}{\delta {F}}\right)_{\!\!\bc}
\de^3c\,,
\\\label{EP-PCS2}
&\frac{\partial}{\partial t}\frac{\delta l}{\delta
\bX}+\pounds_{\bX+\bX_\bU}\,\frac{\delta l}{\delta \bX}=
F\,\nabla_{\bz}\frac{\delta l}{\delta {F}}
\\\label{EP-PCS4}
&\frac{\partial F}{\partial t}+\nabla\cdot\left[(\bX+\bX_\bU) {F}\right]=0
\,.
\end{align}
Here, $\pounds_\bW\boldsymbol{\sigma}=\nabla\cdot(\bW\boldsymbol{\sigma})+\nabla\bW\cdot\boldsymbol{\sigma}$ is the Lie derivative of a one-form density and the subscripts $\bq$ and $\bc$ denote the space and velocity components of vector-valued quantities on phase-space. As shown in \cite{HoTr2011}, a remarkable property of the system \eqref{EP-PCS1}-\eqref{EP-PCS4} is that the equation \eqref{EP-PCS1} can be rewritten equivalently as a Lie transport equation as follows:
\begin{equation}\label{ZLS}
\left(\frac{\partial}{\partial t}+\pounds_{\bU}\right)\!\left(\frac{\delta l}{\delta\bU}-\int\!\frac{\delta l}{\delta \bw}\,\de^3c+\nabla_\bq\cdot\int\!\bc\,\frac{\delta l}{\delta \ba}\,\de^3c\right)=0
\,.
\end{equation}
As we shall see, this relation has important consequences as it allows enforcing the  zero-average condition on velocity fluctuations.

In this section, we shall show how the mean-fluctuation splitting \eqref{EQ1}-\eqref{EQ2} arises from equations \eqref{EP-PCS1}-\eqref{EP-PCS4} by adopting the following Lagrangian, essentially given by the sum of the particle phase-space Lagrangian and the fluid Lagrangian for the mean-flow:
\begin{equation}\label{Lagr1}
l=\int\! F\!\left[\left(m\bc+m\bU+q\bA+\bgamma\right)\cdot\mathbf{w}-\frac{m}2|\bc|^2-q\Phi+\frac{m}2|\bU|^2+q\bU\cdot\bA\right]\de^3x\,\de^3c
\,.
\end{equation}
Here, $(\Phi,\bA)$ are the potentials of the  external (static) electromagnetic field, while $\bgamma(\bq,t)$ is an extra dynamical variable playing the role of a  Lagrange multiplier enforcing the necessary zero-average condition on velocity fluctuations, that is $\int\! F\bw\,\de^3\bc=0$. 
The mean-fluctuation splitting system \eqref{EQ1}-\eqref{EQ2} arises by replacing the functional derivatives
\begin{align*}
\frac{\delta l}{\delta \bX}=&\ F\big(m\bc+m\bU+q\bA+\boldsymbol\gamma,0\big)
\,,
\\
\frac{\delta l}{\delta \bU}=&\ mn\bU+qn\bA
\,,
\\
\frac{\delta l}{\delta F}=&\ \left(m\bc+m\bU+q\bA+\boldsymbol\gamma\right)\cdot\mathbf{w}-\frac{m}2|\bc|^2-q\Phi+\frac{m}2|\bU|^2+q\bU\cdot\bA
\end{align*}
in \eqref{EP-PCS1}-\eqref{EP-PCS4} and by making use of the constraint $\int\!F\bw\,\de^3c=0$. 
As a first step, we notice that, the Lagrange multiplier $\boldsymbol\gamma$ can be set to zero {\color{black} after taking variations}, since equation \eqref{ZLS} implies
\[
\left(\frac{\partial}{\partial t}+\pounds_{\bU}\right)(n\bgamma)=0
\,,
\]
which indeed possesses the trivial solution $\bgamma(\bq,t)=0$. In what follows we shall always select this particular solution {\color{black} after taking variations}. With this in mind, after some vector calculus, equation \eqref{EP-PCS2} leads to
\[
\bw(\bq,\bc,t)=\bc
\,,\qquad\qquad
\ba(\bq,\bc,t)=\frac{q}{m}\bE+\frac{q}{m}(\bc+\bU)\times\bB-{\partial_t\bU}-(2\bc+\bU)\cdot\nabla\bU\,.
\]
Then, making use of this result in equation \eqref{EP-PCS1} yields \eqref{EQ1} and  \eqref{EQ2}.

The advantage of the  treatment presented here lies in the fact that, at this point, the Lagrangian \eqref{Lagr1} may be modified at will, depending on the desired approximation. For example, the assumption of a negligible mean-flow inertia leads to discarding the first two $\bU-$terms in \eqref{Lagr1} and this is exactly the approach followed in \cite{Tronci2014}. In this paper, we shall present a similar application of this approach in the case when the fundamental kinetic equation (before the mean-fluctuation splitting is performed) is given by Littlejohn's drift-kinetic equation of guiding-center motion \cite{CaBr,Littlejohn}. This is precisely the topic of the next section and, as we shall see, the resulting system sheds a new light on different features of guiding-center dynamics.

\section{Remarks on guiding-center motion and drift-kinetic theory}

In this section, we present some features of guiding-center motion that will be relevant for later discussions, especially concerning drift-fluid models and their properties.

\subsection{Alternative variational settings for guiding-center motion}

Before we enter the discussion on the role of the the mean-fluctuation splitting in drift-kinetic theory, it is useful to introduce an alternative variational formulation of guiding-center theory, which differs from Littlejohn's well-known Lagrangian
\begin{equation}\label{LLagr}
L(\bq,\dot\bq,v_\|)=(mv_\|+q\bA(\bq))\cdot\dot\bq-\frac{m}2v_\|^2-\mu B(\bq)-q\Phi(\bq)
\,.
\end{equation}
Notice that here, we simply treat the magnetic moment $\mu$ as a physical constant thereby ignoring the gyrophase. While Littlejohn's Lagrangian \eqref{LLagr} is entirely written on the guiding-center phase-space, an alternative Lagrangian is also available on physical space and this reads as follows:
\begin{equation}\label{Lagr2}
\mathcal{L}(\bq,\dot\bq)=\frac{m}2(\bb(\bq)\cdot\dot\bq)^2+q\dot\bq\cdot\bA(\bq)-\mu B(\bq)-q\Phi(\bq)
\end{equation}
Then, upon denoting  
\begin{equation}
\bE^*=-\nabla\Phi-\frac{\mu}{q}\nabla B
\,,\qquad
\bB^*=\bB+\frac{m}{q}(\bb\cdot\dot\bq)\nabla\times\bb
\,,
\end{equation}
the guiding-center equation emerges as a standard Euler-Lagrange equation on physical space, that is
\[
q\bE^*+q\dot\bq\times\bB^*=m(\bb\cdot\ddot\bq)\bb
\,.
\]
The general idea of carrying the guiding-center position along with the 3-dimensional particle velocity goes back a long time and we address the reader to \cite{Goldston}. 
The Lagrangian \eqref{Lagr2} may be particularly helpful to identify conserved quantities associated to certain symmetries. In the case $\Phi=0$, the simplest example is given by a spatially constant magnetic field,  leading to conservation of the total linear momentum 
\[
\frac{\partial \cal L}{\partial \dot{\bq}}=m(\bb\cdot\dot\bq)\bb+q\bA
\,.
\]
However, in the context of the present paper, the Lagrangian \eqref{Lagr2} is particularly useful to understand the construction underlying the mean-fluctuation splitting. More specifically, since the mean-flow equation is a fluid equation in physical, then a description of guiding-center motion in the configuration space becomes advantageous. Indeed, Lagrangian fluid paths are always defined on configuration space, rather than phase-space.

The Lagrangian \eqref{Lagr2} offers the possibility of lifting guiding-center motion on $\Bbb{R}^4$ to the standard six-dimensional phase-space with corresponding phase-space Lagrangian given by
\begin{equation}\label{Lagr2PS}
{\cal L}'(\bq,\dot\bq,\bv)= \big(m\bb(\bq)\bb(\bq)\cdot \bv+q\bA(\bq)\big)\cdot\dot\bq-\frac{m}2(\bb(\bq)\cdot \bv)^2-\mu B(\bq)-q\Phi(\bq)
\,.
\end{equation}
This Lagrangian is highly degenerate and obviously no information can be obtained on the perpendicular components of the velocity. Nevertheless, lifting guiding-center motion to $\Bbb{R}^6$ is particularly advantageous as a preliminary step for the formulation of hybrid kinetic-fluid descriptions. As we remarked above, this is due to the fact that fluid and particle paths can be composed with each other, thereby restoring the fundamental duality between Lagrangian and Eulerian pictures in continuum mechanics. As an example, this approach based on lifting guiding-center motion to $\Bbb{R}^6$ was recently exploited in \cite{ClBuTr} for the formulation of hybrid MHD models in the pressure-coupling scheme. This is precisely the strategy  adopted later on in this paper.

\subsection{Drift-kinetic momentum density evolution}

In preparation for the sections below, here we shall also present the exact evolution equation for the guiding-center momentum density, which is computed as the first-order moment of Littlejohn's drift-kinetic equation for a static external electromagnetic field. Littlejohn's equation for the guiding-center distribution $f(\bq,v_\|,t)$ is considered here in the following form:
\begin{equation}\label{DK-eq}
\frac{\partial f}{\partial t}+\nabla\cdot( f \bu)+\frac{\partial f}{\partial v_\|}(f\alpha_\|)=0
\,.
\end{equation}
where  $(\bu,\alpha_\|)$ is the phase-space vector field defined by
\begin{equation}\label{X-VF}
q\bE^*+q\bu\times\bB^*=m\alpha_\|\bb
\,,
\end{equation}
so that, upon denoting $B_\|^*=\bb\cdot\bB^*=B+(mv_\|/q)\bb\cdot\nabla\times\bb$, 
\begin{equation}\label{GC-VF}
\bu(\bq,v_\|,t)=\frac1{B_\|^*}\big(v_\|{\bB^*}-{\bb }\times\bE^*\big)
\,,\qquad\quad
\alpha_\|(\bq,v_\|,t)=\frac{q}{mB_\|^*}\, {{\bB^*}\cdot\bE^*}
\,.
\end{equation}
As we shall see, the algebraic relation \eqref{X-VF} may be extremely useful when computing moments of the drift-kinetic equation \eqref{DK-eq}. Indeed, equation \eqref{X-VF} is much more fundamental than its algebraic solution \eqref{GC-VF} for $(\bu,\alpha_\|)$.

At this point, we introduce the notation $\bK=m\int\!f v_\|\bb\,\de v_\|\de\mu$ for the momentum density and we are ready to compute its evolution equation. To this purpose, we write
\begin{align}\nonumber
	\frac1m\frac{\pa\bK}{\pa t}
	&=-\nabla\cdot\int\! f\vp\bu\bb \,\de\vp\de\mu
	+\int \!f\vp\bu\cdot\nabla\bb \,\de\vp\de\mu
	+\int\! f \alpha_\parallel\bb \,\de\vp\de\mu
	\\
	&=-\nabla\cdot\int\! f\vp\bu\bb \,\de\vp\de\mu
	+\int \!f\vp\nabla\bb\cdot\bu \,\de\vp\de\mu
		+\int \!f\big(q\bu\times\bB-\mu\nabla B+q\bE\big)\,\de\vp\de\mu
	\,.
	\label{mario}
\end{align}
Eventually, upon denoting $\bu_\perp=-\bb\times\bb\times\bu$ and by using the relations 
\[
	\int\! f(\vp\nabla\bb\cdot\bu -\mu\nabla B)\,\de\vp\de\mu
	=\nabla\bB\cdot\bM
	=(\nabla\times\bM)\times\bB
	+\nabla\cdot\big[(\bB\cdot\bM)\boldsymbol{1}
	-\bB\bM\big]\,,
\]
we are left with the following evolution equation for the guiding-center momentum density:
\begin{equation}\label{momdens}
\frac{\pa\bK}{\pa t}=-\nabla\cdot\Bbb{T}+qn\bE
	+\big(\bJ+\nabla\times\bM\big)\times\bB
	\,,
\end{equation}
where we have defined the stress tensor, the current, and the magnetization respectively as
\begin{align}\label{stressdef}
&\Bbb{T}(\bq,t)=\int\! f\Big[mv_\|^2\bb\bb
	+m\vp\bu_\perp\bb
	+m\vp\bb\bu_\perp
	+\mu B(\boldsymbol{1}-\bb\bb)\Big]\,\de\vp\de\mu
	\\\label{magnetizationdef}
	&\bM(\bq,t)=-\int\!f\mu\bigg[\bb-\frac{m}{\mu B}{\vp}\bu_\perp \bigg]\,\de\vp\de\mu
	\\\label{currentdef}
	&\bJ(\bq,t)=q\int\!f\bu\,\de\vp\de\mu
	\,.
\end{align}
As we can see, the CGL stress tensor in \eqref{stressdef} is completed  by two nongyrotropic terms, which have already appeared in recent work \cite{BrTr,Graves}. Similarly, the definition \eqref{magnetizationdef} of the magnetization requires the addition of a nongyrotropic moment appeared in  \cite{Evstatiev_2014,BrTr}.  

We notice that the momentum density equation \eqref{momdens}  obtained here is exact in the sense that no approximations were used in its derivation. This momentum density equation stands as a fundamental point of departure especially for the formulation of drift-fluid models, which should always arise from \eqref{momdens} upon adopting an adequate moment closure or truncation for the the stress tensor evolution. Indeed, unless a moment closure or truncation is performed, the whole drift-kinetic equation \eqref{DK-eq}-\eqref{X-VF} must be carried over to close the system. Equivalently, in the next section we shall  perform the mean-fluctuation splitting on Littlejohn's drift-kinetic equation by applying  the variational approach presented earlier in Section \ref{sec:Vlasov}.

\section{Mean-fluctuation splitting for drift-kinetic theory}

The mean-fluctuation Lagrangian \eqref{Lagr1} can be easily adapted to drift-kinetic theory by combining the six-dimensional approach  associated to the Lagrangian \eqref{Lagr2PS} with the guiding-center Lagrangian \eqref{Lagr2} on the physical configuration space. The latter is indeed particularly useful to construct the mean-flow (fluid) terms in the mean-fluctuation Lagrangian, which can now be written as
\begin{multline}
l=\int\! F\!\bigg[\left(m\bb\bb\cdot\bc+m\bb\bb\cdot\bU+q\bA+\bgamma\right)\cdot\mathbf{w}
\\
-
\frac{m}2(\bb\cdot\bc)^2-\mu B-q\Phi+\frac{m}2(\bb\cdot\bU)^2+q\bU\cdot\bA\bigg]\de^3x\,\de^3c\,\de\mu
\label{GCsplittingLagr}
\,.
\end{multline}
Again, we have introduced the fluctuation velocity coordinate $\bc=\bv-\bU$ and we recall the constraint $\int \!F\bw\,\de^3 c\,\de\mu=0$.   Also, we emphasize the distinction between the (relative) guiding-center distribution $\tilde{f}(\bq,c_\|,\mu,t)$ and its extended version $F(\bq,\bc,\mu,t)$.
Then, the mean-fluctuation splitting equations can be obtained by substituting
\begin{align*}
\frac{\delta l}{\delta \bX}=&\ f\big(m\bb\cdot\bc+m\bb\cdot\bU+q\bA+\bgamma,\,{0}\big)
\,,\\
\frac{\delta l}{\delta \bU}=&\  mn\bb\bb\cdot\bU+qn\bA
\,,\\
\frac{\delta l}{\delta F}=&\  \left(m\bb\bb\cdot\bc+m\bb\bb\cdot\bU+q\bA+\bgamma\right)\cdot\mathbf{w}-\frac{m}2(\bb\cdot\bc)^2-\mu B-q\Phi+\frac{m}2(\bb\cdot\bU)^2+q\bU\cdot\bA
\end{align*}
in equations \eqref{EP-PCS1}-\eqref{EP-PCS4}. Before doing that, however, we notice again that the relation \eqref{ZLS} leads to selecting the special solution $\boldsymbol\gamma=0$, which will indeed be used from now on. At this point, with the relations above, equation \eqref{EP-PCS2} yields 
\begin{equation}\label{velrel}
\bb\cdot\bw=\bb\cdot\bc
\end{equation}
as well as, after a vector calculus exercise,
\begin{multline}
m\Big[
\bb\cdot(\ba+\bc\cdot\nabla\bU)+(\bw+\bU)\cdot\nabla\bb\cdot \bc\Big]\bb
-q(\bw+\bU)\times\boldsymbol{\cal B}^*
\\
=-m\bb\partial_t(\bb\cdot\bU)
+q\bE^*-m(\bb\cdot\bc)\nabla(\bb\cdot\bU)-\frac{m}2\nabla(\bb\cdot\bU)^2
\,,
\label{accel}
\end{multline}
where  we have introduced the notation
\begin{align*}
\boldsymbol{\cal B}^*
=&\ 
\bB+({m}/{q})
(\bb\cdot\bc)\nabla\times\bb
+({m}/{q})
\nabla\times[\bb\bb\cdot\bU]
\\
=&\ {\bB}^{*}-({m}/{q})\bb\times\nabla(\bb\cdot\bU)
\,.
\end{align*}
Notice that, while $0=\int \!F\bb\cdot\bw\,\de^3 c\,\de\mu=\int \!F\bb\cdot\bc\,\de^3 c\,\de\mu$, we have $0=\int \!F\bw_\perp\de^3 c\,\de\mu\neq\int \!F\bc_\perp\de^3 c\,\de\mu$ and the latter integral will never be set to zero.

In order to proceed further, here we shall recall a result recently appeared in \cite{ClBuTr} (see Appendix A.2 therein) and relating equation \eqref{EP-PCS4} for $F$ to that for $\tilde{f}(\bq,c_\|,\mu,t)=\int\!F(\bq,\bc,\mu,t)\,\de^2 c_\perp$. Indeed, when the magnetic field is independent of time, it is a general result that \eqref{EP-PCS4} implies the following equation for $\tilde{f}$:
\begin{equation}
\frac{\partial \tilde{f}}{\partial t}+\nabla\cdot\int(\bw+\bU) F\,\de^2 c_\perp
+
\frac{\partial}{\partial c_\|}\int\!\big[\bb\cdot(\ba+\bc\cdot\nabla\bU)+(\bw+\bU)\cdot\nabla\bb\cdot \bc\big] F\,\de^2 c_\perp=0
\,.
\label{Close}
\end{equation}
Thus, after verifying that ${\cal B}^*_\|=B^*_\|$, dotting \eqref{accel} with $\boldsymbol{\cal B}^*$ and crossing with $\bb$ leads to the  guiding-center equation for the relative distribution $\tilde{f}(\bq,c_\|,\mu,t)$, which reads
\begin{multline}\label{RelDKeq}
\frac{\partial \tilde{f}}{\partial t}+\nabla\cdot\left[(c_\|+U_\|)\frac{\tilde{f}\boldsymbol{\cal B}^*}{{\cal B}_\|^*}-\frac{m}q\frac{\tilde{f}\bb }{{\cal B}_\|^*}\times\Big(\frac{q}m\bE^*-(c_\|+U_\|) \nabla U_\|\Big)\right]
\\
+
\frac{\partial}{\partial c_\|}\left[
\frac{\tilde{f}\boldsymbol{\cal B}^*}{{\cal B}_\|^*}\cdot\Big(\frac{q\bE^*}{m}-(c_\|+U_\|)\nabla U_\|\Big)-\tilde{f}\frac{\partial U_\|}{\partial t}\right]=0
\,.
\end{multline}
For later purposes, it is useful to notice that the equation above may be written similarly to \eqref{DK-eq} as a continuity equation on the four-dimensional  phase-space
\begin{equation}\label{altDKeq}
\frac{\partial \tilde{f}}{\partial t}+\nabla\cdot(\boldsymbol{w}\tilde{f})+\frac{\partial }{\partial c_\|}(a_\| \tilde{f})=0
\,,
\end{equation}
where the vector field $(\boldsymbol{w},a_\|)$ is given by the solution of the vector equation
\begin{equation}\label{fundVF}
ma_\|\bb-q(\boldsymbol{w}+\bU)\times\boldsymbol{\cal B}^*
=-m\bb\partial_tU_\|
+q\bE^*-mc_\|\nabla U_\|-\frac{m}2\nabla U_\|^2
.
\end{equation}

We now turn our attention to the guiding-center correspondent of the fluid equation  \eqref{EQ1}, that is the drift-fluid equation corresponding to the mean flow. First, we observe that
\[
\int\!\bc\otimes\left(\pounds_{\bX}\frac{\delta l}{\delta\bX}-
F\nabla_{\bz}\frac{\delta l}{\delta F}\right)_{\!\bc}\de^3c\,\de\mu
=0
\,.
\]
Also, after a vector calculus exercise using the relations $\int\! \bw\,F\,\de^3c\,\de\mu=0$ and $\nabla\bb\cdot\bb=0$, we obtain
\begin{multline*}
\int\!\left(\pounds_{\bX}\frac{\delta l}{\delta\bX}-
F\nabla_{\bz}\frac{\delta l}{\delta F}\right)_{\!\bq}\de^3c\,\de\mu=m\!
\int\!\Big[
\nabla\cdot\left(F(\bb\cdot\bc)(\bb\bb\cdot\bc+\bw_\perp)\bb\right)-F\left(\bb\cdot\bc\right)\nabla\bb\cdot\bw_\perp
\Big]\de^3c\,\de\mu
\\
+\bar\mu\nabla B+qn\nabla\Phi-\frac{mn}2\nabla(\bb\cdot\bU)^2-qn\nabla(\bU\cdot\bA)
\,,
\end{multline*}
where we have defined  $n=\int\! F\,\de^3 c\,\de \mu$  and $\bar\mu(\bq,t)=\int\!F\mu\,\de^3c\,\de\mu$. If we now introduce
\begin{align}\nonumber
\boldsymbol{\cal M}(\bq,t)=&\ -\bar\mu\bb+\frac{m}B\int\!{(\bb\cdot\bc)}\,\bw_\perp F\,\de^3c\,\de\mu
\\
=&\ -\bar\mu\left[\bb-\frac{m}{\bar\mu B}\int\!{c_\|}\boldsymbol{w}_\perp \tilde{f}\,\de c_\|\,\de\mu\right]
\label{magndef}
\end{align}
and  make use of the vector relation $\nabla\bA\cdot\bB+\bA\times\nabla\times\bB=\nabla\cdot[(\bA\cdot\bB)\boldsymbol{1}-\bA\bB]$, we can write
 \begin{align*}
\int\!\left(\pounds_{\bX}\frac{\delta l}{\delta\bX}-
F\nabla_{\bz}\frac{\delta l}{\delta F}\right)_{\!\bq}\de^3c\,\de\mu=&\ 
m\nabla\cdot\int\!
F(\bb\cdot\bc)(\bb\bb\cdot\bc+\bw_\perp)\bb
\,\de^3 c\,\de\mu
\\
&\ 
-\nabla \bB\cdot\boldsymbol{\cal M}-\frac{mn}2\nabla(\bb\cdot\bU)^2-qn\nabla(\bU\cdot\bA)-qn\bE
\\
=&\ 
\nabla\cdot{\cal P}
+\bB\times\nabla\times\boldsymbol{\cal M}
-\frac{mn}2\nabla(\bb\cdot\bU)^2-qn\nabla(\bU\cdot\bA)-qn\bE
\,.
\end{align*}
Here, we have defined the drift-kinetic pressure tensor, recently appeared in \cite{BrTr,Graves}
\begin{align}\nonumber
{\cal P}=&\ m\int\!
F(\bb\cdot\bc)\big[(\bb\cdot\bc)\bb\bb+\bw_\perp\bb+\bb\bw_\perp\big]
\,\de^3 c\,\de\mu
+\bar\mu B(\boldsymbol{1}-\bb\bb)
\\
=&\ m\int\!
\tilde{f} c_\|\big(c_\|\bb\bb+\boldsymbol{w}_\perp\bb+\bb\boldsymbol{w}_\perp\big)
\,\de c_\| \,\de\mu
+\bar\mu B(\boldsymbol{1}-\bb\bb)
\,.\label{pressuredef}
\end{align}
%We remark that, while the integral terms in \eqref{pressuredef} comprise a pressure tensor, the last term retains the character of a stress tensor and this is due to the definition of magnetic moment $\mu=(m/2)v_\perp^2/B=(m/2)|\bc_\perp+\bU_{\!\perp}|^2/B$.
Then,
the drift-fluid equation for the mean flow reads as
\[
m\left(\frac{\partial}{\partial t}+\pounds_{\bU}\right)(nU_\|\bb)=-\nabla\cdot{\cal P} +qn\bE+\frac{mn}2\nabla(\bb\cdot\bU)^2+(qn\bU+\nabla\times\boldsymbol{\cal M})\times\bB
\]
or, equivalently,
\begin{equation}\label{meanfloweq}
mn{\partial_t U_\|}\bb+mn(\bU\cdot\nabla U_\|)\bb=-\nabla\cdot{\cal P}+qn\bE+qn\bU\times{\boldsymbol{B}}^*-\bB\times\nabla\times\boldsymbol{\cal M}
\,.
\end{equation}
Here, we have used the notation $\boldsymbol{B}^*=(\int\!{\boldsymbol{\cal B}}^*F\,\de^3 c)/n=\bB+(m/q)U_\|\nabla\times\bb$. Since $\int \!F\,{\bb\cdot\bc}\,\de^3 c\,\de\mu=0$, we notice that the nongyrotropic second moment $\int \!F(\bb\cdot\bc)\bw_\perp\,\de^3 c\,\de\mu$ contained in $\Bbb{P}$ and $\boldsymbol{\cal M}$ does not depend on $\bU_{\!\perp}$ and thus the velocity vector field $\bU$ can be given an algebraic expression in terms of its parallel component $U_\|$. This is obtained by crossing  equation \eqref{meanfloweq} with $\bb$ so that
\begin{equation}\label{U-eq}
\bU=U_\|\frac{{\boldsymbol{B}}^*}{{{B}}^*_\|}+\frac{\bb}{qn{{B}}^*_\|}\times(\bB\times\nabla\times\boldsymbol{\cal M}+\nabla\cdot{\cal P}-qn\bE)
\,.
\end{equation}
On the other hand, dotting equation \eqref{meanfloweq} with $\boldsymbol{\cal B}^*$ yields
\begin{equation}\label{meanfloweq-bis}
{\partial_t U_\|}+\bU\cdot\nabla U_\|=-\frac1{mnB^*_\|}\, \boldsymbol{B}^*\cdot\big(\nabla\cdot{\cal P}+\bB\times\nabla\times\boldsymbol{\cal M}-qn\bE
\big)
\,.
\end{equation}
Equations \eqref{meanfloweq-bis}, \eqref{U-eq}, and \eqref{RelDKeq} represent the mean-fluctuation splitting of drift-kinetic equations, which are accompanied by the definitions \eqref{magndef} and \eqref{pressuredef}. 

It can be shown that the mean-flow equation \eqref{meanfloweq} is equivalent to the guiding-center momentum equation  \eqref{momdens} found previously. To see this, one starts by verifying that  ${\bw}(\bq,c_\|,t)=\bu(\bq,c_\|+U_\|(\bq,t),t)-\bU(\bq,t)$, where $\bu(\bq,v_\|,t)$ is given as in \eqref{GC-VF}. Indeed, we have
\begin{align*}
\bw(\bq,c_\|,t)=&\ 
(c_\|+U_\|)\frac{\boldsymbol{\cal B}^*}{B_\|^*}-\frac{m}q\frac{\tilde{f}\bb }{{\cal B}_\|^*}\times\Big(\frac{q}m\bE^*-(c_\|+U_\|) \nabla U_\|\Big)-\bU
\\
=&\ 
(c_\|+U_\|)\frac{\bB^*}{B_\|^*}-\frac{\bb }{B_\|^*}\times\bE^*-\bU
\\
=&\ 
\bu(\bq,c_\|+U_\|,t)-\bU
\,.
\end{align*}
Then, upon denoting $\tilde{f}(\bq,c_\|,t)={f}(\bq,c_\|+U_\|,t)$, we rewrite the first two terms on the right hand side of \eqref{mario} as
\begin{align*}
-\nabla\cdot\int \!f\vp\bu\bb^T \de\vp\,\de\mu
	+\int\! f\vp\nabla\bb\cdot\bu \,\de\vp\,\de\mu
=&\ 
-\int\! \bigg\{\nabla\cdot\Big[\tilde{f}(c_\|+U_\|)(\bw\bb^T+\bU\bb^T) \Big]
\\&\ \qquad \quad
	+\tilde{f}(c_\|+U_\|)\nabla\bb\cdot(\bw+\bU) \bigg\}\,\de c_\|\,\de\mu
		\\
	=&\ 
-\nabla\cdot\int \!\tilde{f}\Big[c_\|\bw\bb^T +c_\|\nabla\bb\cdot\bw\Big] \de c_\|\,\de\mu
\\&\ 
	-\pounds_\bU(nU_\|\bb)
	+\frac{n}2\nabla U_\|^2
	\,,
\end{align*}
so that standard vector identities recover \eqref{meanfloweq}.

The mean-fluctuation system comprised by the equations \eqref{meanfloweq} and \eqref{RelDKeq} stands as a fundamental step for the derivation of reduced models  usually obtained from moment truncations or closures so that the accompanying kinetic equation \eqref{RelDKeq} can be replaced by appropriate equations for the other moments. Usually, one neglects moments of order higher than two as well as non-gyrotropic second-order moments and this is the approach presented in the next section.

%%%%%%%%%%%%%%%%%%%%%%%%%%%%%%%%%%%%

\subsection{Gyrotropic drift-fluid models \label{DFmods}}

Starting from the mean-fluctuation system,  second-order moment truncations may be obtained upon recalling  the form \eqref{altDKeq}-\eqref{fundVF} of the drift-kinetic equation \eqref{RelDKeq} and by enforcing the gyrotropic condition
\begin{equation}\label{gyrotropicass}
\int \!\tilde{f}\,c_\|\boldsymbol{w}_\perp\,\de c_\|\,\de\mu=0
\end{equation}
within the moment equations,  so that the pressure tensor \eqref{pressuredef} reduces to the usual CGL form. Also, in this case the magnetization  \eqref{magndef} depends uniquely on the magnetic moment density thereby recovering the familiar-looking relation $\boldsymbol{\cal M}=-\bar\mu \bb$. The next step is the characterization of the magnetic moment density in terms of its equation of motion; it is easy to see that, upon neglecting the magnetic moment flux $\int\! \tilde{f} c_\|\mu\,\de c_\|\,\de \mu$, the magnetic moment density satisfies the continuity equation 
\begin{equation}\label{magmomtransp}
\partial_t\bar\mu+\nabla\cdot(\bar\mu\bU)=0
\,. 
\end{equation}
Notice that this is the same equation satisfied by the particle number, that is  the zero-th moment equation
\begin{equation}\label{denstransp}
\partial_t n+\nabla\cdot(n\bU)=0
\,.
\end{equation}

At this point, one is led to dealing with the longitudinal pressure $p_\|=m\int\!\tilde{f} c_\|^2\,\de c_\|\de\mu$. In the simplest case, the pressure is ignored by the cold-plasma closure, which resorts to the  solution ansatz 
\begin{equation}\label{CFans}
\tilde{f}(\bq,c_\|,\mu,t)=n(\bq,t)\,\delta(c_\|)\,\delta\big(\mu-{\bar\mu(\bq,t)}/{n(\bq,t)}\big)
\,.
\end{equation}
Notice that this ansatz represents a fluid that is `cold' only in the longitudinal direction, while a finite perpendicular temperature is incorporated in the magnetic moment density. For example, this approach was followed in \cite{StSc} in the context of gyrofluids, although the equations presented therein are different from those obtained here. The \emph{cold drift-fluid} ansatz \eqref{CFans} produces the following mean-flow equation
\begin{equation}
mn{\partial_t U_\|}\bb+mn(\bU\cdot\nabla U_\|)\bb=qn{\bE}-\bar\mu\nabla B+qn\bU\times{\boldsymbol{B}}^*.
\label{CFeq}
\end{equation}
In \cite{StSc}, the force $-\bar\mu\nabla B$ associated to the magnetic drift is replaced by a gradient pressure term, thereby making the resulting drift-fluid model similar to those previously proposed in \cite{Pfirsch}. Here, we notice that the cold drift-fluid equation \eqref{CFeq} possesses a variational structure similar to that in \cite{StScBr}. More specifically, equation \eqref{CFeq} arises as the Euler-Poincar\'e equation for the variational principle
\begin{equation}\label{CFvarprin}
\delta\int^{t_2}_{t_1}\!\int\!\left(\frac{mn}2 (\bb\cdot \bU)^2+qn\bU\cdot\bA - \bar\mu B-qn\Phi\right)\de t=0
\end{equation}
with the variations $\delta\bU=\partial_t\boldsymbol\Xi+[\bU,\boldsymbol\Xi]$, $\delta n=-\nabla\cdot( n\boldsymbol\Xi)$ and $\delta \bar\mu=-\nabla\cdot( \bar\mu\boldsymbol\Xi)$. A similar variational structure will be used in later sections. Before proceeding further,  it may be useful to remind the reader that the cold-plasma model stands as a moment \emph{closure} as opposed to a moment \emph{truncation}. Indeed, equation \eqref{CFeq} follows directly from the solution ansatz \eqref{CFans} without any further assumptions. On the contrary, moment truncations are obtained by simply discarding higher-order moment contributions as shown below.

A more sophisticated way of dealing with the longitudinal pressure is by applying the moment method, that is by computing the evolution equation for $p_\|$ from \eqref{RelDKeq} and then truncating to discard heat-flux contributions. To this purpose, we multiply \eqref{RelDKeq} by $c_\|\bb\bb$ and integrate over the parallel velocity before taking the trace. Upon denoting $q_\|=m\int\!\tilde{f} c_\|^3\,\de c_\|\de\mu$, this proceeds as follows:
\begin{align*}
\bb\bb\frac{\partial}{\partial t}\int\!c_\|^2 \tilde{f}\,\de c_\|\de\mu=
& -\bb\bb\nabla\cdot\int\!c_\|^2(\boldsymbol{w}+\bU)\tilde{f}\,\de c_\|\de\mu
\\
&\ 
+2\bb\int\!c_\|\left[\frac{q}m(\boldsymbol{w}+\bU)\times\bB^*
-\bb\partial_tU_\|
-\frac{\mu}m\nabla B-c_\|\nabla U_\|-\frac12\nabla U_\|^2\right]\tilde{f}\,\de c_\|\de\mu
\\
=
& -\bb\bb\nabla\cdot\left(p_\|\bU+q_\|\bb\right)
-\bb\bb\nabla\cdot\int\!c_\|^2\boldsymbol{w}_\perp \tilde{f}\,\de c_\|\de\mu
\\
&\ 
+2\bb\int\!c_\|\left[\boldsymbol{w}_\perp\times\left(q\bB+
\nabla\times(c_\|\bb+U_\|\bb)\right)
-\frac{\mu}m\nabla B\right]\tilde{f}\,\de c_\|\de\mu
\\
&\ +\frac{2}m\bb\Big[q_\|\bb\times\nabla\times\bb+p_\|\Big(\bU\times\nabla\times\bb+\bb\times\nabla\times(U_\|\bb)-\nabla U_\|\Big)\Big]
,
\end{align*}
where the first equality follows from writing \eqref{RelDKeq}  as in \eqref{altDKeq}-\eqref{fundVF}, while the second equality follows after writing $\boldsymbol{w}=\boldsymbol{w}_\perp+c_\|\bb$ and expanding. Then, taking the trace and invoking \eqref{gyrotropicass} leads to 
\begin{equation}
\frac{\partial p_\|}{\partial t}= -\nabla\cdot(p_\|\bU+q_\|\bb+\boldsymbol{q}_\perp)-2
\bb\bb:(p_\|\nabla\bU+\nabla\boldsymbol{q}_\perp)+2q_\mu\nabla\cdot\bb
\,,
\label{Macmahoneq2}
\end{equation}
where $\boldsymbol{q}_\perp=m\int\!\tilde{f}c_\|^2\bw_\perp \de c_\|\de\mu$ and $q_\mu=B\int\!\tilde{f}c_\|\mu\, \de c_\|\de\mu$. {\color{black}Here, the special case $\boldsymbol{q}_\perp=0$ consistently recovers equation (16) in \cite{Hammett}. In turn,} we notice that equation \eqref{Macmahoneq2} differs quite substantially from the celebrated parallel CGL equation \cite{CGL,MondtWeiland}. Indeed, while \eqref{Macmahoneq2} has been obtained here exclusively from guiding-center theory, the original CGL derivation \cite{CGL,MondtWeiland} relies on special solutions of the truncated equation for the full-orbit pressure tensor dynamics. Nevertheless, the parallel CGL equation coincides   with \eqref{Macmahoneq2} when heat flux contributions are neglected, so that
\begin{equation}
\frac{\partial p_\|}{\partial t}+\bU\cdot\nabla p_\|= -p_\|\nabla\cdot\bU-2
p_\|\bb\bb:\nabla\bU
\,.
\label{Macmahoneq}
\end{equation}
In the context of MHD models, due to the frozen-in condition on the magnetic field, equation \eqref{Macmahoneq} also implies an adiabatic invariant and it can be obtained from a thermodynamic equation of state \cite{Hazeltine,HoKu}, upon letting the internal energy depend on the magnetic field. As pointed out in \cite{Brizard}, the frozen-in condition on the magnetic field is strictly necessary for the emergence of an adiabatic invariant associated to the parallel pressure. Here,  we notice that \eqref{Macmahoneq} can be written in terms of the Lie derivative of the tensor density $p_\|b_ib_j\,\de x^i \de x^j\otimes\de^3 x$ as follows:
\[
\operatorname{Tr}\left[\bigg(\frac{\partial}{\partial t}+\pounds_\bU\bigg)(p_\|\bb\bb)\right]=0
\,,
\]
where $\operatorname{Tr}$ denotes the matrix trace and in this case the Lie derivative reads $(\pounds_\bU S)_{ij}={\bU\cdot\nabla S_{ij}}+ (\nabla\cdot\bU)S_{ij}+S_{kj}\partial_iU^k + S_{ki}\partial_jU^k$. Equation \eqref{Macmahoneq} accompanies the drift-fluid equation, which extends the cold-fluid model \eqref{CFeq} as follows:
\begin{align}\nonumber
mn{\partial_t U_\|}\bb+mn(\bU\cdot\nabla U_\|)\bb=&\, -\nabla\cdot(p_\| \bb\bb)+qn\boldsymbol{E}^*+qn\bU\times{\boldsymbol{B}}^*
\\
=&\, -\nabla\cdot[p_\| \bb\bb+\bar\mu B(\boldsymbol{1}-\bb\bb)]+qn\bE+qn\bU\times{\boldsymbol{B}}^*-\bB\times\nabla\times\boldsymbol{M},
\label{gyrofmom}
\end{align}
where $\boldsymbol{M}=-\bar\mu\bb$. The moment model \eqref{gyrofmom}-\eqref{Macmahoneq} emerges here as a slight extension of a model previously appeared in \cite{StScBr} and lacking the magnetization term. In turn, this magnetization term appears to be necessary for momentum conservation. The dynamics of the total drift-fluid momentum is actually the topic of the next section.

A third way to deal with pressure in drift-fluid models was first proposed by Pfirsch and Correa-Restrepo (PCR) \cite{Pfirsch} and it consists in replacing the kinetic energy term ${m}c_\|^2/2+\mu B$ in \eqref{GCsplittingLagr} by a fluid internal energy. The resulting fluid model can be obtained from the variational principle \eqref{CFvarprin} upon subtracting the internal energy term $mn\mathcal{U}(n)$. Here we shall follow Pfirsch's approach by adopting a barotropic fluid closure, while adiabatic extensions can be easily formulated to include entropy transport. Then, the following variant of PCR variational principle
\begin{equation}\label{CFvarprin2}
\delta\int^{t_2}_{t_1}\!\int\!\left(\frac{mn}2 (\bb\cdot \bU)^2+qn\bU\cdot\bA -qn\Phi-mn\mathcal{U}(n)\right)\de t=0
\end{equation}
produces the fluid equation 
\begin{equation}
mn{\partial_t U_\|}\bb+mn(\bU\cdot\nabla U_\|)\bb=qn{\bE}+qn\bU\times{\boldsymbol{B}}^*-\nabla p,
\label{CFeq2}
\end{equation}
where $p=mn\mathcal{U}'(n)$.  While Pfirsch obtained an analogous result by including the polarization drift in the variational principle \eqref{CFvarprin2},  equation \eqref{CFeq2} coincides with PCR   when those polarization effects are ignored (see equation (44) in \cite{Pfirsch}).  Also, notice that the magnetic drift arising from  the presence of the magnetic moment density in \eqref{CFeq} is actually absent in \eqref{CFeq2}. Instead, the magnetic drift has been replaced by the diamagnetic drift associated to the isotropic pressure; see Section 7.2 in \cite{Goldston}  for a discussion on how the coexistence of magnetic and  diamagnetic drift is actually excluded by physical arguments.

\subsection{Drift-fluid models and hydrodynamic helicity\label{sec:hel}}
In this section, we want to study the implications of the drift-fluid momentum equation  in different cases. We begin the discussion by considering equation \eqref{gyrofmom} in the equivalent Lie-derivative form
\[
\left(\frac{\partial}{\partial t}+\pounds_{\bU}\right)(qn\bA+mnU_\|\bb)=-\nabla\cdot(p_\| \bb\bb)-\bar\mu\nabla B+\frac{mn}2\nabla(\bb\cdot\bU)^2+qn\nabla(\bU\cdot\bA)
\,,
\]
thereby leading to the following explicit equation for circulation dynamics:
\[
\frac{\de}{\de t}\oint_{\Gamma(t)}\left[\frac{n}{\bar{\mu}}(q\bA+mU_\|\bb)\right]\cdot\de\bq=\oint_{\Gamma(t)}\frac1{\bar\mu}\left[\frac{mn}2\nabla(\bb\cdot\bU)^2+qn\nabla(\bU\cdot\bA)-\nabla\cdot(p_\| \bb\bb)\right]\cdot\de\bq
\,, 
\]
for an arbitrary loop $\Gamma(t)$ moving with velocity $\bU$. Since $\Gamma(t)$ is arbitrary, the above circulation equation leads to the vorticity evolution by applying Stokes theorem. Thus, the terms generating circulation are also terms generating vorticity.

Let us first consider the case of a cold drift-fluid, that is $p_\|=0$. Unlike the full-orbit version of the cold-fluid equation, we notice that circulation is generated by the gradients of the ratio $n/\bar\mu$, which then represent a source of magnetic reconnection that is absent in standard full-orbit fluid models. If we insist on full consistency with the full-orbit case, then reconnection sources can be eliminated by restricting to consider a constant ratio $n/\bar\mu$, which indeed is allowed by the equation $\partial_t(n/\bar\mu)+\bU\cdot\nabla(n/\bar\mu)=0$. Notice that this is in agreement with the barotropic limit (isentropic) of equation (47) in \cite{Hazeltine}, where the authors propose defining a characteristic perpendicular length such that $d_\perp=\bar\mu/n=const.$

Now let us consider the case in which $\bar\mu/n=d_\perp$ and the longitudinal pressure is retained, so that the circulation law becomes
\[
\frac{\de}{\de t}\oint_{\Gamma(t)}(q\bA+mU_\|\bb)\cdot\de\bq=-d_\perp\oint_{\Gamma(t)}\bigg[\frac1{\bar\mu}\nabla\cdot(p_\| \bb\bb)\bigg]\cdot\de\bq
\,.
\]
As we can see, circulation is now generated by the longitudinal pressure. For the sake of completeness, we should remark that in this case  \eqref{gyrofmom} produces the following equation of motion for the hydrodynamic helicity:
\[
\frac{\de}{\de t}\int(q\bA+mU_\|\bb)\cdot(q\bB+m\nabla\times(U_\|\bb))\,\de^3x=-2 d_\perp\int(q\bB+m\nabla\times(U_\|\bb))\cdot\bigg[\frac1{\bar\mu}\nabla\cdot(p_\| \bb\bb)\bigg]\de^3 x
\]
Thus, the  longitudinal pressure stands as another source of vorticity. We notice that this is very different from the case of isentropic full-orbit fluid models, whose pressure forces do not generate vorticity.  Now, since the circulation generated by the longitudinal pressure also represents a source of magnetic reconnection, one has to be careful about the physical effects incorporated in the model. For example, the Hall-MHD model does not comprise magnetic reconnection in the non-resistive case. 
%Thus, we conclude that it is not possible to consider drift-fluid models as approximations of isentropic full-orbit fluid equations, unless one goes back to the cold-plasma equation \eqref{CFeq} and adopts $\bar\mu=d_\perp n$. 

A possible way to formulate drift-fluid models whose pressure forces mimic the full-orbit case is found in the PCR approach \cite{Pfirsch}. Indeed, in this case, equation \eqref{CFeq2} preserves the circulation integral $\oint_{\,\Gamma}(q\bA+mU_\|\bb)\cdot\de\bq$, thereby leading to helicity conservation 
\[
\frac{\de}{\de t}\int({q}\bA+mU_\|\bb)\cdot(q\bB+{m}\nabla\times(U_\|\bb))\,\de^3x=qm\frac{\de}{\de t}\int (B+B^*_\|)U_\|\,\de^3x=0
\]
and reproducing the analogous result from full-orbit models. Thus, in this sense, the PCR drift-fluid method leads to drift-fluid models that successfully reproduce guiding-center drifts while retaining the invariants from the full-orbit theory.

{\color{black}\subsection{Gyrotropic equations of state\label{sec:CGLeqst}}
As pointed out in Section \ref{DFmods}, the existence of a longitudinal adiabatic invariant $\partial_t(p_\|B^2/n^3)+\bU\cdot\nabla(p_\|B^2/n^3)=0$ arises in CGL fluid theory \cite{CGL} since in that particular case the magnetic field is frozen-in. In the general case, no longitudinal adiabatic invariant can be shown to emerge from standard guiding-center theory. On the other hand, one may be tempted to enforce the existence of such an invariant by enforcing a \emph{CGL equation of state} of the type
\begin{equation}\label{CGLeqst}
%\frac{p_\|B^2}{n^3}=m^3h(s)
%\quad\implies\quad
p_\|=\frac{m^3n^3}{B^2}h(s)
%=:\rho\,\mathcal{U}(\rho,\tilde\mu,B)-\frac\rho{m}\tilde\mu B
\,,
\end{equation}
where $h$ is some fixed real function and $s({\bf x},t)$ is a transported scalar satisfying $\partial_ts+ \bU\cdot\nabla s=0$. For example, this scalar can be the specific entropy \cite{HoKu}, although here we shall chose the specific magnetic moment $s=\bar\mu/n$. Then, since $p_\|=m\int\!\tilde{f} c_\|^2\,\de c_\|\de\mu$, the guiding-center kinetic energy associated to velocity fluctuations can be rewritten as
\begin{equation}\label{int-en}
\int\!\tilde{f}\left(\frac{m}2c_\|^2+\mu B\right)\de c_\|\,\de\mu=mn\,
\mathcal{U}(n,s)=mn\left(\frac{m^2n^2}{2B^2}h(s)+\frac{s}m B\right),
\end{equation}
thereby leading to the following anisotropic modification of the PCR variational principle \eqref{CFvarprin2}:
\begin{equation}\label{CGLeqstvarpr}
\delta\int^{t_2}_{t_1}\!\int\!\left(\frac{mn}2 (\bb\cdot \bU)^2+qn\bU\cdot\bA -qn\Phi-mn\mathcal{U}(n,s)\right)\de t=0
\end{equation}
Here, the variations are again given by $\delta\bU=\partial_t\boldsymbol\Xi+[\bU,\boldsymbol\Xi]$,  $\delta n=-\nabla\cdot( n\boldsymbol\Xi)$, and $\delta \bar\mu=-\nabla\cdot( \bar\mu\boldsymbol\Xi)$. Thus, we compute
\[
\delta p_\|=\delta\bigg(\frac{m^3n^3}{B^2}h(s)\bigg)=  -\nabla\cdot(p_\|\boldsymbol\Xi)-2p_\|\bb\bb:\nabla\boldsymbol\Xi+2\frac{p_\|}B\bb\cdot\nabla\times(\boldsymbol\Xi\times\bB)
\,,
\]
where the last term appears by writing  $0=\delta\bB=\nabla\times(\boldsymbol\Xi\times\bB)-\nabla\times(\boldsymbol\Xi\times\bB)$, since we are dealing with an external electromagnetic field. The mean flow equation reads as
\begin{multline}
mn{\partial_t U_\|}\bb+mn(\bU\cdot\nabla U_\|)\bb
=qn\bE+qn\bU\times{\boldsymbol{B}}^*
\\
 -\nabla\cdot[p_\| \bb\bb+\bar\mu B(\boldsymbol{1}-\bb\bb)]+\bB\times\nabla\times((\bar\mu+p_\|/B)\bb).
\label{gyrofmom2}
\end{multline}
Thus, the last term shows that enforcing the longitudinal adiabatic invariant in guiding-center motion produces an extra magnetization current as the magnetization density is now redefined to include parallel pressure contributions, i.e. $\boldsymbol{M}=-(p_\|/B+\bar\mu)\bb$. In addition, we observe that if $\bar\mu=d_\perp n$ as in Section \ref{sec:hel}, then the drift-fluid model  \eqref{denstransp}-\eqref{gyrofmom2} preserves Kelvin's circulation $\oint_{\,\Gamma(t)}(q\bA+mU_\|\bb)\cdot\de\bq$ as well as the hydrodynamic helicity $\int (B+B^*_\|)U_\|\,\de^3x$. Thus, while invoking extra magnetization effects, the CGL equation of state recovers consistency with full-orbit models in terms of circulation conservation and so it does not comprise possible sources of magnetic reconnection. The same arguments hold for possible extensions of the CGL equation of state that have previously been proposed in \cite{Daughton}. Therein, the parallel pressure $p_\|$ is expressed as a more general function ${\cal F}_\|$ of the ratio $n^3/B^2$, so that $p_\|={\cal F}_\|(n^3/B^2)$. Analogously, one defines two functions ${\cal F}_{\!\perp}^{\,(1)}$ and ${\cal F}_{\!\perp}^{\,(2)}$ so that the magnetic moment density is expressed as $\bar\mu =n{\cal F}_{\!\perp}^{\,(1)}(n^3/B^2)+nB^{-1}{\cal F}_{\!\perp}^{\,(2)}(n^3/B^2)$. Then, one can still define a (barotropic) fluid internal energy ${\cal U}(n)$ by an appropriate modification of \eqref{int-en}, thereby preserving both circulation and hydrodynamic helicity.
}

\section{The mean-fluctuation splitting as a modeling framework}

In this section, we explain how the mean-fluctuation splitting may serve as a modeling framework for models that carry physical features from both full-orbit picture and the guiding-center picture. This specific possibility is offered by the fact that, while the mean and the fluctuation velocities in \eqref{Lagr1} and \eqref{GCsplittingLagr} are treated on the same footing, different descriptions may be allowed depending on modeling purposes. For example, here we shall consider two possible options. Starting from the Lagrangian functional \eqref{GCsplittingLagr}, we shall consider two intermediate models in which 1) fluctuations are treated as in the full-orbit theory, while the mean-flow is described by a drift-fluid Lagrangian and 2) fluctuations are given as in the guiding-center description, while the mean flow is associated to the usual charged fluid Lagrangian. As we shall see, the first variant provides a basis for the PCR pressure closure of gyrofluids, while the second variant leads to fluid models incorporating the magnetic drift within the full-orbit description.

\subsection{Full-orbit fluctuations and guiding-center mean flow\label{PfirschMetod}}

In the first model we consider, the fluctuation terms in the Lagrangian  \eqref{GCsplittingLagr} are replaced by the full-orbit fluctuation terms appearing in \eqref{Lagr1}. This step produces the following modified Lagrangian
\begin{multline}
l=\int\! F\bigg[\left(m\bc+m\bb\bb\cdot\bU+q\bA+\bgamma\right)\cdot\mathbf{w}
-
\frac{m}2c^2-q\Phi+\frac{m}2(\bb\cdot\bU)^2+q\bU\cdot\bA\bigg]\de^3x\,\de^3c
\label{GCsplittingLagr2}
\,,
\end{multline}
where the distribution function is $F=F(\bq,\bc,t)$. The resulting equations of motion are again obtained by specializing \eqref{EP-PCS1}-\eqref{EP-PCS4}. In this case, we obtain $\bw=\bc$ and
\begin{equation*}
m(
\ba+\bc\cdot\nabla\bU)
-(\bc+\bU)\times\big[q\bB+m \nabla\times(U_\|\bb)\big]
=-m\bb\partial_tU_\|
+q\bE-m\nabla\bU\cdot \bc-\frac{m}2\nabla U_\|^2
\,.
%\label{accel2}
\end{equation*}
Then, upon denoting
\[
\bE^\star=\bE-({m}/q)\nabla\bU\cdot \bc
\,,\qquad
\bB^\star=\bB+({m}/q) \nabla\times(U_\|\bb)
\,,
\]
and after noticing that $\nabla_\bz\cdot(\bX+\bX_\bU)=0$ (see the notation in Section \ref{sec:Vlasov}),
the kinetic equation reads
\begin{align}\nonumber
\frac{\partial F}{\partial t}+(\bc+\bU)\cdot\frac{\partial F}{\partial \bq}+\Big[(q/m)(\bc+\bU)\times\bB^\star
+(q/m)\bE^\star-\bb\partial_tU_\|-U_\|\nabla U_\|\Big]\cdot\frac{\partial F}{\partial \bc}=0
\,.
\end{align}
In addition, since $\left(\pounds_{\bX}({\delta l}/{\delta\bX})-
F\nabla_{\bz}({\delta l}/{\delta F})\right)_{\bc}=0$ and 
\[
\int\!\left(\pounds_{\bX}\frac{\delta l}{\delta\bX}-
F\nabla_{\bz}\frac{\delta l}{\delta F}\right)_{\!\bq\!}\de^3c=-\nabla\cdot\Bbb{P}-\frac{mn}2\nabla(\bb\cdot\bU)^2-qn\nabla(\bU\cdot\bA)-qn\bE
\,,
\]
the fluid equation \eqref{EP-PCS1} reads
\begin{equation}\label{meanfloweq2}
mn{\partial_t U_\|}\bb+mn(\bU\cdot\nabla U_\|)\bb=-\nabla\cdot\Bbb{P}+qn\bE+qn\bU\times{\boldsymbol{B}}^*
\,,
\end{equation}
with $\boldsymbol{B}^*=\bB+(m/q)U_\|\nabla\times\bb$ and we recall $\Bbb{P}=\int\!F\bc\bc\,\de^3 c$. We notice that, upon introducing the hat map notation $\widehat{A}_{ij}=-\epsilon_{ijk} A_k$, the pressure tensor evolution reads
\begin{equation}\label{FOpressEvol}
\frac{\partial{\Bbb{P}}}{\partial t}+(\bU\cdot\nabla){\Bbb{P}}+(\nabla\cdot\bU){\Bbb{P}}+{\Bbb{P}}\cdot\nabla\bU+\big({\Bbb{P}}\cdot\nabla\bU\big)^T
+\left[({q}/{m})\widehat{B}^*-\widehat{\omega}\,,\,{\Bbb{P}}\right]
=-\nabla\cdot\boldsymbol{\mathsf{Q}}
\,,
\end{equation}
where $[\cdot,\cdot]$ denotes the matrix commutator, $\boldsymbol\omega=\nabla\times\bU$ is the mean flow vorticity, and $\boldsymbol{\mathsf{Q}}=\int\!F\bc\bc\bc\,\de^3 c$ is the heat flux tensor. It is easy to see that this equation possesses gyrotropic solutions recovering the CGL evolution equations \cite{CGL,MondtWeiland} and thus this approach recovers the gyrofluid model in \cite{StScBr}. In addition, at this point one can  adopt an isotropic  equation of state as in the standard approach to fluid dynamics; this  leads naturally to equation \eqref{CFeq2} as in the PCR model \cite{Pfirsch} discussed previously.  {\color{black}Alternatively, retaining non-gyrotropic  components of the full pressure tensor can be advantageous, for example, in capturing magnetic reconnection effects while still exploiting the simplifications offered by the use of the guiding-center approximation to describe the mean-fluid flow fluid.}

%Then, here we have shown that letting fluctuations evolve according to full-orbit dynamics, while adopting the guiding-center description for the mean flow, allows naturally for an isotropic pressure in gyrofluid models, which  instead cannot be obtained from the full guiding-center treatment.

\subsection{Guiding-center fluctuations and full-orbit mean flow}
In this section, we shall consider the alternative approach to that described above. That is, here we shall treat the mean fluid flow as a standard full-orbit charged fluid, while we shall retain the guiding-center description to describe fluctuation dynamics. This leads immediately to the following Lagrangian:
\begin{multline}
l=\int\! F\bigg[\left(m\bb\bb\cdot\bc+m\bU+q\bA+\bgamma\right)\cdot\mathbf{w}
\\
-
\frac{m}2(\bb\cdot\bc)^2-\mu B-q\Phi+\frac{m}2U^2+q\bU\cdot\bA\bigg]\de^3x\,\de^3c\,\de\mu
\label{GCsplittingLagr3}
\,.
\end{multline}
Again, the equations of motion are obtained by specializing \eqref{EP-PCS1}-\eqref{EP-PCS4}. In this case, we obtain $\bb\cdot\bw=\bb\cdot\bc$ as well as, after a vector calculus exercise,
\begin{multline*}
m\Big[
\bb\cdot(\ba+\bc\cdot\nabla\bU)+(\bw+\bU)\cdot\nabla\bb\cdot \bc\Big]\bb
-q(\bw+\bU)\times\boldsymbol{\cal B}^\star
\\
=-m\partial_t\bU
+q\bE^*-m(\bb\cdot\bc)\nabla(\bb\cdot\bU)-\frac{m}2\nabla U^2
\,,
\end{multline*}
where we have used
\[
\boldsymbol{\cal B}^\star=\bB+({m}/{q})
(\bb\cdot\bc)\nabla\times\bb+({m}/q)\nabla\times\bU\,.
\]
Then, upon using \eqref{Close}, we have the kinetic equation
\begin{multline}\label{RelDKeq2}
\frac{\partial \tilde{f}}{\partial t}+\nabla\cdot\left[(c_\|+U_\|)\frac{\tilde{f}\boldsymbol{\cal B}^\star}{{\cal B}_\|^\star}-\frac{m}q\frac{\tilde{f}\bb }{{\cal B}_\|^\star}\times\bigg(\frac{q\bE^*}{m}-c_\|\nabla U_\|- \nabla U^2/2-\frac{\partial \bU}{\partial t}\bigg)\right]
\\
+
\frac{\partial}{\partial c_\|}\left[
\frac{\tilde{f}\boldsymbol{\cal B}^\star}{{\cal B}_\|^\star}\cdot\bigg(\frac{q\bE^*}{m}-c_\|\nabla U_\|- \nabla U^2/2-\frac{\partial \bU}{\partial t}\bigg)\right]=0
\,.
\end{multline}
This can be equivalently written in the form \eqref{altDKeq}, where the vector field $(\boldsymbol{w},a_\|)$ is now given by the solution of the algebraic equation
\begin{equation}\label{fundVF2}
ma_\|\bb-q(\boldsymbol{w}+\bU)\times\boldsymbol{\cal B}^\star
=-m\partial_t\bU
+q\bE^*-mc_\|\nabla U_\|-\frac{m}2\nabla U^2
\,.
\end{equation}
Upon taking the second moment and by neglecting the heat flux contributions, it is easy to see that this formulation of the kinetic equation \eqref{RelDKeq2} recovers the CGL equation \eqref{Macmahoneq} for the parallel pressure $p_\|$.

In order to write the accompanying fluid equation, we first verify that $\left(\pounds_{\bX}({\delta l}/{\delta\bX})\right)_{\bc}=
F\nabla_{\bc}({\delta l}/{\delta F})$ and then we compute
\begin{align*}
\int\!\left(\pounds_{\bX}\frac{\delta l}{\delta\bX}-
F\nabla_{\bz}\frac{\delta l}{\delta F}\right)_{\!\bq}\de^3c\,\de\mu
%=&\ m\!\int\!\Big[\nabla\cdot\left(F(\bb\cdot\bc)(\bb\bb\cdot\bc+\bw_\perp)\bb\right)-F\left(\bb\cdot\bc\right)\nabla\bb\cdot\bw_\perp\Big]\de^3c\,\de\mu
%\\&\ 
%+\bar\mu\nabla B+qn\nabla\Phi-\frac{mn}2\nabla U^2-qn\nabla(\bU\cdot\bA)
%\\
=&\ 
\nabla\cdot{\cal P}
+\bB\times\nabla\times\boldsymbol{\cal M}
-\frac{mn}2\nabla U^2-qn\nabla(\bU\cdot\bA)-qn\bE
\,,
\end{align*}
with the same notation as in \eqref{magndef} and \eqref{pressuredef}. Thus, equation 
\eqref{EP-PCS1} yields the mean-flow equation in the form
\begin{equation}\label{meanfloweq3}
mn\left({\partial_t \bU}+\bU\cdot\nabla \bU\right)=-\nabla\cdot{\cal P}+qn\bE+\left(qn\bU+\nabla\times\boldsymbol{\cal M}\right)\times{\bB}
%-\bB\times\nabla\times\boldsymbol{\cal M}
\,.
\end{equation}
We recall that this equation is accompanied by the transport equations \eqref{magmomtransp} and \eqref{denstransp} for  the magnetic moment density  and the particle density, respectively.
Thus, we notice that, even after enforcing a gyrotropic truncation so that $\int \!c_\|\boldsymbol{w}_\perp\tilde{f}\,\de c_\|\de\mu=0$, the guiding-center motion of fluctuation dynamics manifests in the mean-flow equation through the magnetic drift and the parallel pressure contribution. Indeed, in this case equation \eqref{meanfloweq3} leads to the gyrofluid model given by the equation
\begin{equation}\label{meanfloweq3bis}
mn\left({\partial_t \bU}+\bU\cdot\nabla \bU\right)=qn\bE+qn\bU\times{\bB}-\nabla\cdot(p_\|\bb\bb)-\bar\mu\nabla B
%-\bB\times\nabla\times\boldsymbol{\cal M}
\,,
\end{equation}
which is itself accompanied by the parallel CGL closure  \eqref{Macmahoneq}, as well as \eqref{magmomtransp} and \eqref{denstransp}. This suggests that, when fluctuations are goverend by guiding-center motion, anisotropic fluid models based on the CGL pressure tensor $p_\|\bb+p_\perp(\boldsymbol{1}-\bb\bb)$ should actually be accompanied by a magnetization term appearing in the perpendicular pressure  and arising from  the relation 
\begin{align*}
-\bar\mu\nabla B=\nabla\bB\cdot\boldsymbol{M}&
=-\bB\times\nabla\times\boldsymbol{M}+\nabla\cdot\left[(\bB\cdot\boldsymbol{M})\boldsymbol{1}-\bB\boldsymbol{M}\right]
\\
&=
\bB\times\nabla\times(\bar\mu\bb)-\nabla\cdot\big(\bar\mu B(\boldsymbol{1}-\bb\bb)\big)
\,.
\end{align*}
However, the magnetization term cancels in the presence of a frozen-in condition for the magnetic field \cite{Hazeltine,HoKu} and this is the reason why magnetization effects do not appear explicitly in the original CGL model.

%The magnetization term did not emerge in the original work \cite{Chew} because, as mentioned earlier, the CGL closure was derived from gyrotropic solutions of the full-orbit pressure equation and thus in this treatment the perpendicular pressure was not at all related to the properties of the magnetic moment in guiding-center theory. Nevertheless, the perpendicular adiabatic invariant is still present in the original CGL theory, since the latter invokes the frozen-in condition of the magnetic field that is typical in MHD-type models. These observations lead naturally to the possibility of extending the original CGL model to incorporate the guiding-center magnetization. This is the subject of the following section.

These last two sections showed how the mean-fluctuation splitting leads to different modeling options depending on the description adopted for the mean flow and the fluctuation dynamics. In the next sections, we shall go back to standard drift-fluid models in which both the mean flow and fluctuations are described by guiding-center theory and we shall illustrate the level of difficulty of these models when they are coupled to the Maxwell equations for the self-consistent evolution of the electromagnetic field.

%%%%%%%%%%%%%%%%%%%%%%%%%%%

\section{Maxwell's equations and the role of magnetization}
In  this section, we extend the previous discussions to the case of dynamic electromagnetic fields obeying Maxwell's equations. 
The coupling of guiding-center theory with Maxwell's equations was first considered by Pfirsch \cite{Pfirsch1} and was recently developed further in \cite{BrTr}, while the numerical discretizations of the resulting models are found in \cite{Evstatiev_2014,Eero}.

 The equations of the drift-kinetic Maxwell system are written as follows: the kinetic equation is the same as in \eqref{DK-eq}-\eqref{X-VF}, upon replacing
\begin{equation}\label{newEF}
\bE^*=-\frac{\partial \bA}{\partial t}-\nabla\Phi-\frac{\mu}q\nabla B-\frac{m}qv_\|\frac{\partial \bb}{\partial t}\,,
\end{equation}
while Amp\`ere's law reads
\begin{equation}\label{DKAmpere}
\mu_0^{-1}\nabla\times\bB-\varepsilon_0\frac{\partial \bE}{\partial t}=\bJ+\nabla\times{\bM}
\,,
\end{equation}
with the definitions \eqref{magnetizationdef}-\eqref{currentdef}. On the other hand, Gauss' law and Faraday's law remain unchanged as in standard electromagnetism. In the next sections, we shall reformulate this model by applying the mean-fluctuation splitting from the previous sections. Eventually, drift-fluid models will be obtained by applying the methods in Section \ref{DFmods}.

\subsection{Mean-fluctuation splitting\label{mark1}}
In the presence of dynamic electromagnetic fields, the mean-fluctuation splitting Lagrangian \eqref{GCsplittingLagr} is added to the Maxwell Lagrangian
\begin{equation}
l_{\scriptscriptstyle{\rm Max}}=\frac12\int\!\bigg(\varepsilon_0\left|\nabla\Phi+\frac{\partial \bA}{\partial t}\right|^2-\frac1{\mu_0}|\nabla\times\bA|^2\bigg)\,\de^3x
\,,
\end{equation}
so that the new Lagrangian is 
\begin{equation}
\ell=l+l_{\scriptscriptstyle{\rm Max}}
\,,
\label{Lagr3}
\end{equation} 
where $l$ is given by \eqref{GCsplittingLagr}. {\color{black}We take the occasion to emphasize that, consistently with Littlejohn's theory, the Lagrangian \eqref{Lagr3} assumes an $E\times B$ speed much smaller than the particle thermal speed \cite{CaBr}. Alternatively, should these speeds be comparable, we would simply subtract the $E\times B$ energy term $m_h B^{-2}\left|\bE\times\bb\right|^{2}\!/2$ from the guiding center kinetic energy $m_hv_\|^2/2+\mu B$, thereby producing extra magnetization terms along with polarization effects \cite{Krommes2} which are instead ignored here. Then, as a result of the Euler-Poincar\'e variational principle $\delta\int_{t_1}^{t_2}\ell\,\de t=0$,}
the equations \eqref{EP-PCS1}-\eqref{EP-PCS4} do not change in form, although the time-dependent magnetic field now appears in \eqref{EP-PCS2} so that \eqref{accel}  becomes 
\begin{multline*}
m\Big[
\bb\cdot(\ba+\bc\cdot\nabla\bU)+(\bu+\bU)\cdot\nabla\bb\cdot \bc+\frac{\partial \bb}{\partial t}\cdot\bc\Big]\bb
-q(\bw+\bU)\times\boldsymbol{\cal B}^*
\\
=-m\bb\partial_t(\bb\cdot\bU)
+q\bE^*-m(\bb\cdot\bc)\nabla(\bb\cdot\bU)-m\frac12\nabla(\bb\cdot\bU)^2
\,,
%\label{accel2}
\end{multline*}
where
\begin{equation}\label{newElField}
\bE^*=-\frac{\partial \bA}{\partial t}-\nabla\Phi-(\mu/q)\nabla B-(m/q)(\bb\cdot\bc+\bb\cdot\bU)\frac{\partial \bb}{\partial t}
\,.
\end{equation}
Now, in the case of a time-dependent magnetic field, the relation \eqref{Close} is modified as \cite{ClBuTr}
\begin{equation}
\frac{\partial \tilde{f}}{\partial t}+\nabla\cdot\int(\bw+\bU) F\,\de^2 c_\perp
+
\frac{\partial}{\partial c_\|}\int\!\big[\bb\cdot(\ba+\bc\cdot\nabla\bU)+(\bw+\bU)\cdot\nabla\bb\cdot \bc+\bc\cdot\partial_t\bb\big] F\,\de^2 c_\perp=0
\,.
\label{Close2}
\end{equation}
and since  \eqref{velrel} remains unchanged, one obtains formally the same drift-kinetic equation \eqref{RelDKeq}, although with the definition \eqref{newElField}  in the form $\bE^*=\bE-(\mu/q)\nabla B-(m/q)(c_\|+U_\|){\partial_t \bb}$. Notice that this drift-kinetic equation can be again written as in \eqref{altDKeq}-\eqref{fundVF}.

As a next step, we turn our attention to the fluid equation \eqref{EP-PCS1}. Since the functional derivatives of the Lagrangian $\ell$ in \eqref{Lagr3} with respect to $(\bU,\bX,F)$ are the same as the functional derivatives of the Lagrangian $l$ in \eqref{GCsplittingLagr} with respect to $(\bU,\bX,{F})$, equation \eqref{EP-PCS1} produces the following variant of the  drift-fluid equation \eqref{meanfloweq}:
\begin{equation}\label{meanfloweq2bis}
mn({\partial_t U_\|}+\bU\cdot\nabla U_\|)\bb=-\nabla\cdot{\cal P}+qn\bE-mnU_\|\partial_t\bb+qn\bU\times{\boldsymbol{B}}^*-\bB\times\nabla\times\boldsymbol{\cal M}
\,,
\end{equation}
where $\boldsymbol{\cal M}(\bq,t)$ and ${\cal P}(\bq,t)$  are given by  \eqref{magndef} and  \eqref{pressuredef}, respectively, upon using \eqref{newElField} in the relation \eqref{fundVF} defining the vector field $\boldsymbol{w}(\bq,c_\|,t)$. Notice that the electric field and its gradients appear on the right-hand side of \eqref{meanfloweq2bis} in three different instances: while the first appearance is explicit, the second is through $\partial_t\bb$ (by Faraday's law, $\partial_t\bB=-\nabla\times\bE$), and the third is through the vector field $\boldsymbol{w}$ itself, which is contained in the magnetization $\boldsymbol{\cal M}$. Again, since the perpendicular component of the velocity $\bU_{\!\perp}$ is absent in $\boldsymbol{\cal M}(\bq,t)$, $\Bbb{P}(\bq,t)$, and $\partial_t\bb$, this allows finding an explicit algebraic expression for the velocity and equation \eqref{U-eq} is modified as follows:
\begin{equation}\label{U-eq2}
\bU=U_\|\frac{{\boldsymbol{B}}^*}{{{B}}^*_\|}+\frac{\bb}{mn{{B}}^*_\|}\times\Big(\bB\times\nabla\times\boldsymbol{\cal M}+\nabla\cdot\Bbb{P}-qn\bE+mnU_\|\partial_t\bb\Big)
\,.
\end{equation}

At this point, we are ready to write down Maxwell's equations by taking variations of $\ell$ in \eqref{Lagr3} with respect to $\Phi$ and $\bA$. While the first yield the usual Gauss law $\nabla\cdot\bE=qn$, the second leads to the following Amp\`ere's law
\begin{equation}\label{Ampere}
\mu_0^{-1}\nabla\times\bB-\varepsilon_0\frac{\partial \bE}{\partial t}=qn\bU+\nabla\times\bigg(\boldsymbol{\cal M}+\frac{mnU_\|}{B}\,\bU_{\!\perp}\bigg)
\,.
\end{equation}
We notice that the term in parenthesis coincides with the full magnetization \eqref{magnetizationdef} appearing in the momentum equation \eqref{momdens}. Upon recalling the vector field $\boldsymbol{w}$ in \eqref{fundVF}, with the definition \eqref{newElField} in the  form $\bE^*=\bE-(\mu/q)\nabla B-(m/q)(c_\|+U_\|){\partial_t \bb}$, this fact is verified by substituting $v_\|=c_\|+U_\|$ and $\bu=\boldsymbol{w}+\bU$ in \eqref{magnetizationdef}. Here, it is useful to explicitly adapt equation \eqref{magndef}  to the present case, that is
\begin{align}\nonumber
\boldsymbol{\cal M}(\bq,t)
=&\ -\bar\mu\bb+\frac{m}B\int\!{c_\|}\bigg[(c_\|+U_\|)\frac{{f}{\bB}^*_\perp}{{ B}_\|^*}-\frac{{f}\bb }{{ B}_\|^*}\times\Big(\bE-\mu\nabla B-(c_\|+U_\|){\partial_t \bb}\Big)\bigg] \,\de c_\|\,\de\mu
\,,
\label{magndef2}
\end{align}
where $\bB^*=\bB+(m/q)(c_\|+\bU_\|)\nabla\times\bb$.
This shows that   gradients  of the electric field up to third order appear in the right-hand side of Amp\`ere's law \eqref{Ampere} through  $\boldsymbol{\cal M}$ and $\bU_{\!\perp}$, thereby making this equation extremely challenging in practice.

Finally, we conclude that applying the mean-fluctuation splitting method to the guiding-center-Maxwell system leads to the equations of motion \eqref{RelDKeq}, \eqref{meanfloweq2bis}-\eqref{U-eq2}, and \eqref{Ampere}. These are accompanied by the definitions \eqref{magndef} and \eqref{pressuredef}, where the vector field $\boldsymbol{w}$ satisfies \eqref{fundVF} upon using $\bE^*=\bE-(\mu/q)\nabla B-(m/q)(c_\|+U_\|){\partial_t \bb}$. However, as remarked above, this system is particularly challenging because of the third-order gradients of the electric field occurring in Amp\`ere's law \eqref{Ampere}.  This problem is partially addressed in the fluid approximation, for which only second order gradients appear in the current balance. This is presented in the next section.

\subsection{Gyrotropic drift-fluid-Maxwell systems}
Upon enforcing the gyrotropic approximation \eqref{gyrotropicass}, the mean-flow equation \eqref{meanfloweq2bis} reduces to 
\begin{equation}\label{meanfloweq3tris}
mn({\partial_t U_\|}+\bU\cdot\nabla U_\|)\bb=-\nabla \cdot(p_\|\bb\bb)+qn\bE-\bar\mu\nabla B+qn\bU\times{\boldsymbol{B}}^*-mnU_\|\partial_t\bb
\,.
\end{equation}
Already at this stage, upon replacing Faraday's law in the last term, we see that, unlike equation \eqref{meanfloweq2}, only first-order gradients of the electric field appear on the right-hand side. Then, for consistency, one applies \eqref{gyrotropicass} to Amp\`ere's law \eqref{Ampere} thereby obtaining
\begin{equation}
\mu_0^{-1}\nabla\times\bB-\varepsilon_0\frac{\partial \bE}{\partial t}=
qn\bU-\nabla\times\bigg(\bar\mu\bb-\frac{mnU_\|}{B}\,\bU_{\!\perp}\bigg)
\,,
\label{Ampere2}
\end{equation}
where the fluid velocity is found from \eqref{meanfloweq3tris} as follows:
\[
\bU=U_\|\frac{{\boldsymbol{B}}^*}{{{B}}^*_\|}+\frac{\bb}{mn{{B}}^*_\|}\times\Big(\nabla \cdot(p_\|\bb\bb)+\bar\mu\nabla B-qn\bE+\frac{mnU_\|}{B}(\nabla\times\bE)_\perp\Big)
\,.
\]
Then, we notice that Amp\`ere's law \eqref{Ampere2} now involves up to second-order gradients of the magnetic field, thereby providing a simplification over the full mean-fluctuation system. It is easy to see that analogous second-order gradient terms also appear in the current balance \eqref{DKAmpere} of the full drift-kinetic Maxwell system. 

In the present setting, an auxiliary equation must be retained for the parallel pressure and this can be provided by the CGL equation \eqref{Macmahoneq}, which is indeed unaffected by the coupling to Maxwell's equations. Unless one resorts to the cold plasma closure \eqref{CFans}, a further possibility to deal with the pressure terms in \eqref{meanfloweq3tris} is  provided by the PCR approach, which replaces the parallel pressure and magnetic moment terms by the gradient of an isotropic pressure associated to an internal energy. {\color{black}Alternatively, one can resort to the CGL equation of state \eqref{CGLeqst} from Section \ref{sec:CGLeqst}, so that the internal energy in \eqref{int-en} now depends also on the magnetic field, i.e. ${\cal U}={\cal U}(n,s,B)$. In this case, one considers the variational principle determined by the sum of the Maxwell Lagrangian $l_{\scriptscriptstyle{\rm Max}}$ and the Lagrangian in \eqref{CGLeqstvarpr}. The, equation \eqref{gyrofmom3} is modified as follows:
\begin{multline}
mn({\partial_t U_\|}+\bU\cdot\nabla U_\|)\bb
=qn\bE+qn\bU\times{\boldsymbol{B}}^*-mnU_\|\partial_t\bb
\\
 -\nabla\cdot[p_\| \bb\bb+\bar\mu B(\boldsymbol{1}-\bb\bb)]+\bB\times\nabla\times((\bar\mu+p_\|/B)\bb), 
\label{gyrofmom3}
\end{multline}
while Amp\`ere's law becomes 
\begin{equation}
\mu_0^{-1}\nabla\times\bB-\varepsilon_0\frac{\partial \bE}{\partial t}=
qn\bU-\nabla\times\bigg[\frac1B\Big({p_\|}\bb+\bar\mu \bB-{mnU_\|}\bU_{\!\perp}\Big)\bigg]
\,.
\label{Ampere3}
\end{equation}
Once again, we observe how the enforcement of the longitudinal adiabatic invariant through the CGL equation of state leads to redefining the magnetization current via the replacement $\bar\mu\to\bar\mu+p_\|/B$ within the magnetization density vector. Also, we emphasize that the coupling to Maxwell's equations leaves preservation of Kelvin's circulation entirely unaffected thereby leading again to the conservation of the hydrodynamic helicity.
}

\section{Getting around the magnetization problem}

The discussions in the previous section have shown how challenging drift-fluid models become when guiding-center motion is coupled to Maxwell equations. Indeed, this is due to the emergence of magnetization terms that introduce higher order gradients in  Amp\`ere's current balance. These considerations lead to the conclusion that drift-fluid models retaining the self-consistent evolution of the electromagnetic field are prohibitively challenging. Nevertheless, at the level of the full kinetic theory, different versions of the drift-kinetic Maxwell model have been discretized in \cite{Evstatiev_2014,Eero}.

In this section, we propose a strategy for the formulation of drift-fluid models that can possibly overcome the difficulties arising from magnetization currents. Here, this is achieved by revisiting the standard theory of guiding-center motion in two separate steps. First, we shall decompose the magnetic field into a fixed (possibly inhomogeneous) background magnetic field and a time-dependent part whose modulus is considerably smaller (although still finite) than the background. Second, we shall perform a guiding-center approximation to average out the fast rotations around the background magnetic field, which is always supposed to be much higher than the gyro-frequency associated to the fluctuating part. The advantage of this approach is that it avoids the emergence of extra magnetization currents other than those normally associated to the magnetic moment. In the following section, we briefly describe the formulation of the theory, which is then followed by its applications to hybrid kinetic-fluid models.

\subsection{Revisiting guiding-center theory\label{sec:modGC}}

In this section, we present a revisitation of guiding-center theory that averages out the fast gyromotion around a high-intensity background magnetic field. We emphasize that this is only one among several possible approaches and, in the present case, the main advantage is probably the level of simplicity of the resulting theory.

As a first step, we decompose the total magnetic field into a background and a time-dependent part, that is
\[
\bB(\bq,t)=\bB_0(\bq)+\widetilde{\bB}(\bq,t)
\,.
\]
Notice that we $\nabla\cdot\bB(\bq,t)=0$ enforces $\nabla\cdot\widetilde{\bB}(\bq,t)=-\nabla\cdot\bB_0(\bq)$.
Here, the modified guiding-center theory is obtained by resorting to a treatment recently appeared in \cite{Tronci2016}, based on kinematic relations and averaged Lagrangians. Although closely related to Littlejohn's approach in \cite{Littlejohn}, the present method does not invoke any previous knowledge on the gyroradius or the magnetic moment. 
As customary in guiding-center theory, we start the discussion with the phase-space Lagrangian in the form
\begin{equation}\label{LilLagr}
L(\br,\dot\br)=(\epsilon\bv+\bA(\br,t))\cdot\dot\br-\frac\epsilon2 v^2
\,,
\end{equation}
where $\bB_0(\bq)+\widetilde{\bB}(\bq,t)=\nabla\times\bA(\bq,t)$ and $\epsilon$ is the smallness parameter. Notice that here we have omitted the electric potential, which can always be restored without essential changes in the final result. Following the treatment in \cite{Tronci2016}, here we decompose the particle position as follows
\begin{equation}
\br(t)=\bX(t)+\epsilon\rho(t)\ba(\bX(t),t)
\,,
%\rho(t) R(\theta(t),\bq)\ba_0
\end{equation}
where
\begin{equation}
\ba(\bq,t)= R(\bq,t)\mathbf{e}_1(\bq)
\,.
\end{equation}
While $\mathbf{e}_1(\bq)$ is an arbitrary field such that $\bb_0\times\mathbf{e}_1=0$, the rotation matrix $R$ is constructed as a rotation around $\bb_0$ by an angle $\theta$ and it is written explicitly as
\[
R(\theta(t),\bb_0(\bq))=\boldsymbol{1}+\sin \theta(t)\,\widehat{\bb}_0(\bq)+(1-\cos\theta(t))\,\widehat{\bb}_0(\bq)\widehat{\bb}_0(\bq)
\,,
\]
where we  write $\theta=\epsilon^{-1}\Theta$ and we  use the hat map introduced in \eqref{FOpressEvol}. Further details on these definitions and the following calculations are found in \cite{Tronci2016}. Notice that here we are restricting to consider a gyroradius that is always perpendicular to the background magnetic field, while in the conventional approach the gyroradius is perpendicular to the total magnetic field. 
As customary in guiding-center motion, we expand the magnetic potential as $\bA(\br,t)=\bA(\bX,t)+\epsilon\rho\ba(\bX,t)\cdot\nabla\bA(\bX,t)$ to write
\begin{equation}\label{aux1}
\bA(\br,t)\cdot\dot\br=\bA(\bX,t)\cdot\dot\bX+\frac\epsilon2 \rho^2\dot\Theta\bb_0(\bX)\cdot \bB(\bX,t)+ \mathcal{G}+O(\epsilon^2)
\,.
\end{equation}
Here, $\mathcal{G}$ denotes the sum of terms that are either total time derivatives or linear in $\ba(\bX,t)$. The total time derivatives in the Lagrangian are well known to be irrelevant to the resulting dynamics. Likewise, the terms linear in $\ba(\bX,t)$ vanish under the final averaging process over $\theta$ and are therefore of no interest for the present purpose. As a further step, we use an orthonormal basis $(\bb_0,\ba,\boldsymbol{h})$ to write $\bv=v_\|\bb_0+n\ba+w\boldsymbol{h}$, so that 
\begin{equation}\label{aux2}
\epsilon\bv\cdot\dot\br=\epsilon\bv\cdot(\dot\bX+\rho\dot\Theta\bb_0(\bX)\times\ba(\bX,t))+O(\epsilon)=\epsilon\bv\cdot\dot\bX-\epsilon\rho\dot\Theta w+O(\epsilon)
\,.
\end{equation}
Finally, replacing \eqref{aux1} and \eqref{aux2} in \eqref{LilLagr} and averaging over $\theta$ leads to the following Lagrangian
\[
\langle L\rangle=(v_\|\bb_0+\epsilon^{-1}\bA)\cdot\dot\bX+\left(\frac12\epsilon^{-1}\rho_L B_\| - w\right)\rho_L\dot\theta-\frac12(v_\|^2+w^2+n^2)
\,,
\]
where we have introduced the Larmor radius $\rho_L=\epsilon\rho$ and we have defined $B_\|=\bb_0\cdot\bB$. Thus, taking variations yields $n=0$ along with
\begin{equation}\label{ModGCRel}
\rho_L=\frac{\epsilon w}{B_\|}
\,,\qquad\quad
\dot\theta=-\epsilon^{-1} B_\|
\,,\qquad\quad
\rho_L^2 B_\|=const.%:\frac{2\epsilon }{q}\mu_\|
\end{equation}
We notice that here the gyration frequency associated to the gyromotion around $\bb_0$  is actually time-dependent as it comprises contributions from the time-dependent magnetic field, due to the presence of $B_\|=B_0+(\bb_0\cdot\tilde\bb)\widetilde{B}$ in the denominator. 

At this point, the above Lagrangian can be cast in a more familiar form upon using the first relation in \eqref{ModGCRel} and introducing the parallel magnetic moment $\mu_\|={mw^2}/({2B_\|})$. Upon restoring the electric potential, one obtains
\begin{equation}
L=(mv_\|\bb_0+q\bA)\cdot\dot\bX-\epsilon\mu_\|\dot\theta-\frac{m}2v_\|^2 - \mu_\| \bb_0\cdot\bB-q\Phi
\,.
\end{equation}
This  is a modification of Littlejohn's Lagrangian in the sense that the time-dependent magnetic potential $\bA$ appears only in the magnetic moment term and in the minimal coupling term, while all other terms involve only the fixed background field $\bB_0$. The resulting equations of motion read
\[
q\bE^\dagger+q\dot{\bX}\times\bB^\dagger=m\dot{v}_\|\bb_0
\,,
\]
where we have defined
\[
\bB^\dagger%=\bB^*_0+\widetilde{\bB}
=\bB+({m}/q)v_\|\nabla\times\bb_0
\,,\qquad
\bE^\dagger=\bE-(\mu_\|/q)\nabla B_\|
\,.
\]
The explicit equations of motion can then be derived as usual by using standard vector algebra. Then, upon defining $B_\|^\dagger=B_\|+({m}/q)v_\|\bb_0\cdot\nabla\times\bb_0$, the corresponding kinetic equation reads
\begin{equation}
\frac{\partial F}{\partial t}+\frac{\partial}{\partial \bq}\cdot\Bigg[\frac{F}{B_\|^\dagger}\big(v_\|{\bB^\dagger}-{\bb_0 }\times\bE^\dagger\big)\Bigg]+\frac{q}{m}\frac{\partial}{\partial v_\|}\Bigg(\frac{F}{B_\|^\dagger}\, {{\bB^\dagger}\cdot\bE^\dagger}\Bigg)=0
\,.
\label{modGCKinEq}
\end{equation}
At this point, the mean-fluctuation splitting may be applied thereby returning an alternative class of gyrofluid models to those presented previously. In this case, the gyrotropic condition \eqref{gyrotropicass} (here, gyrotropic means with respect to $\bb_0$), it is easy to see that, upon neglecting heat-flux contributions, the parallel pressure obeys equation \eqref{Macmahoneq} with $\bb$ replaced by $\bb_0$. Here, one can resort to the CGL equation of state from Section \ref{sec:CGLeqst} (upon replacing $B\to B_0$) or, alternatively, to a non-gyrotropic pressure by following the approach in  Section \ref{PfirschMetod}.

While the pressure tensor arising from the present theory is non-gyrotropic, one advantage of this model is the fact that magnetization currents do not lead to the difficulties mentioned in Section \ref{mark1}. Indeed, upon denoting $\bar\mu_\| =\int\!F\mu_\|\,\de v\de\mu_\|$,  Amp\`ere's law corresponding to the present guiding-center model reads as follows:
\begin{equation}
\mu_0^{-1}\nabla\times\bB-\varepsilon_0\frac{\partial \bE}{\partial t}=
qn\bU-\nabla\times(\bar\mu_\|\bb_0)
\,,
\label{Ampere3}
\end{equation}
which represents a substantial simplification to equation \eqref{Ampere}. A similar Amp\`ere law is obtained by enforcing the CGL equation of state $p_\|=h(s){m^3n^3}/{B_0^2}$, except in this case one replaces $\bar\mu_\|\to p_\|/B+\bar\mu_\|$ in \eqref{Ampere3}. Then, the present model offers the opportunity for drift-ordered models carrying the self consistent evolution of the electromagnetic field while avoiding the difficulties encountered in the standard guiding-center theory.
In the next section, we shall present an application of this guiding-center model to a hybrid kinetic-fluid scheme for energetic particle effects.

%%%%%%%%%%%%%%%%%%%%%%%%%%%%%%%%%%%%%%

\subsection{A hybrid kinetic-drift MHD model}

This section presents an application of the modified guiding-center theory beyond drift-fluid models. More specifically, we shall follow the variational approach in \cite{BuTr} to formulate a hybrid kinetic-drift MHD model in the current-coupling scheme (CCS) aimed at capturing energetic particle effects in fusion devices. Further details are presented in \cite{BuTr,ChSuTr,HoTr2011,PaBeFuTaStSu,ToSaWaWaHo,Tronci2010,TrTaCaMo}, to which we refer the reader for more discussions on hybrid MHD models for energetic particles. {\color{black}Hybrid MHD models describe coupled systems comprising a MHD fluid bulk that interacts with an ensemble of energetic alpha particles obeying kinetic theory (drift-kinetic in this case). These energetic particles are naturally generated in fusion reactions and their effects have been studied over several decades}. 
Motivated by the variational principles presented in \cite{BuTr,HoTr2011}, here we shall construct a hybrid model starting from the following variational principle
\begin{multline}
\delta\int_{t_1}^{t_2}\!\bigg[\int\! F\bigg(\left(mv_\|\bb_0+q\bA\right)\cdot\boldsymbol{u}
-
\frac{m}2v_\|^2-\mu_\| B_\|-q\bU\cdot\bA\bigg)\de^3x\,\de v_\|\,\de\mu
\\
+\int\!\left(\frac{\rho}2U^2-\rho\mathcal{U}(\rho)-\frac1{2\mu_0} B^2\right)\,\de^3 x\bigg]\,\de t=0
\label{CCSVP}
\,.
\end{multline}
Here, the variations are 
\[
\delta\bU=\partial_t\boldsymbol\Xi+[\bU,\boldsymbol\Xi]
\,,\qquad\quad
\delta \rho=-\nabla\cdot( \rho\boldsymbol\Xi)
\] 
for $\rho$ and $\bU$, while we also have 
\[
\delta\bA=\boldsymbol\Xi\times\bB-\nabla(\boldsymbol\Xi\cdot\bA)
\,,\qquad
\delta{\bX}=\partial_t\mathbf{Y}+[{\bX},{\mathbf{Y}}]
\,,\qquad
\delta F=-\nabla_{\bz}(F\mathbf{Y} )
\,.
\]
Here,  $\bX(\bq,v_\|)=(\boldsymbol{u}(\bq,v_\|),a(\bq,v_\|))$ is the guiding-center vector field such that $\partial_t F+\nabla\cdot(\boldsymbol{u} F)+\partial_{v_\|}(a F)=0$ and $\mathbf{Y}$ is again a phase-space displacement vector field analogous to $\boldsymbol\Xi$. The resulting hybrid model is formed of equation \eqref{modGCKinEq} with $\bE^\dagger=-\bU\times\bB-(\mu_\|/q)\nabla B_\|$, and the two advection equations 
\begin{equation}\label{auxiliary}
\partial_t \bB+\nabla\times(\bB\times\bU)=0
\,,\qquad\qquad
\partial_t \rho+\nabla\cdot( \rho\bU)=0
\,.
\end{equation}
Also, we obtain the following MHD momentum equation
\begin{equation}\label{CCSeq}
\rho\partial_t\bU+\rho\left(\bU\cdot\nabla\right)\bU
=\Big(\mu_0^{-1}\nabla\times\bB+qn\bU-\bJ_{\rm gc}-\nabla\times{\bf M}_{\rm gc}\Big)\times\bB-\nabla{ p}
\,,
\end{equation}
where  ${\bf M}_{\rm gc}=-\int\!F\mu_\|\bb_0\,\de v_\|\de\mu_\|$ and 
$
\bJ_{\rm gc}=q\int\!{F}(v_\|\bB^\dagger-\bb_0\times\bE^\dagger)/{B^\dagger_\|}\,\de v_\|\de\mu_\|
$. 

While this equation has a similar structure to the corresponding momentum equation in the Hamiltonian CCS from \cite{BuTr} (see equations (35) and (37) therein), we notice that substantial simplifications occur in the present case. In first instance, no time derivative of the magnetic field orientation occurs in the effective electric field $\bE^\dagger$. Also, and perhaps more importantly, the magnetization term in \eqref{CCSeq} does not involve moments of the kinetic distribution other than the magnetization density. This last aspect establishes an analogy between the present (simplified) Hamiltonian CCS model and the non-Hamiltonian variant implemented in the MEGA code \cite{ToSaWaWaHo}. Indeed, one of the main differences between the Hamiltonian CCS in \cite{BuTr} and its non-Hamiltonian counterpart lies indeed in the fact that the former is substantially more involved due to the presence of extra magnetization terms involving higher velocity moments of the guiding-center distribution. This difficulty is removed in the present variant, which also conserves the standard cross helicity $\int\!\bB\cdot\bU\,\de^3 x$. {\color{black}As noticed in \cite{BuTr}, the momentum equation underlying the MEGA code breaks energy conservation and this is mainly due to the fact that the magnetization density defined in the MEGA code involves the time-dependent magnetic field direction $\bb(\bq,t)$. Indeed, the presence of the entire time-dependent magnetic field in the magnetization density requires the addition of the extra terms in \eqref{magnetizationdef}, which instead are ignored in the MEGA code \cite{ToSaWaWaHo}. On the other hand, in the case of the model \eqref{CCSeq}-\eqref{modGCKinEq} energy conservation is still ensured by the variational structure \eqref{CCSVP}, while the definition of the magnetization density is substantially simplified.}

%%%%%%%%%%%%%%%%%%%%%%%%%%%%%%%%%%%%%%

\section{Conclusions}

This paper has exploited the variational structure of the mean-fluctuation splitting in kinetic theory to shed a new light on the formulation of drift-ordered fluid models for magnetized plasmas. As shown, the application of Euler-Poincar\'e theory to the mean-fluctuation splitting in guiding-center dynamics requires embedding the 4-dimensional guiding-center phase-space into the ordinary 6-dimensional phase-space in order to allow the coupling between fluid paths in configuration space and the phase-space paths arising in kinetic theory. As a result, different types of drift-fluid models were obtained depending on the closure or truncation adopted for the longitudinal pressure. Different truncations and closure variants were considered and previous gyrotropic models \cite{StSc,StScBr} were corrected by the insertion of a magnetization term in the fluid momentum equation. In addition, it was shown that the parallel pressure emerges as a source of magnetic reconnection destroying the cross-helicity conservation characterizing full-orbit fluid models.

As a further step, this paper showed how the mean-fluctuation splitting may actually be used to adopt different physical descriptions for the mean flow and fluctuation kinetics, the particular choice depending on modeling purposes. Two possible options were considered: in the first, the mean-flow is governed by a drift-ordered fluid, while fluctuation kinetics obeys the Vlasov description; in the second case, the converse scenario was presented. 

Special attention was also addressed to the coupling between guiding-center theory (and its drift-fluid closures) and Maxwell's equations for the evolution of the electromagnetic field. In this case, the emergence of specific magnetization correction terms in Amp\`ere's current balance leads to particularly challenging aspects as third-order gradients of the electric field appear in the equations. In order to simplify the treatment, a modified guiding-center model was proposed along with its application to hybrid current-coupling schemes for energetic particle effects.

Among several open questions emerging from this work, we mention the possibility of isolating a Hamiltonian structure for the drift-fluid models presented here. Indeed, while this is easy to find in the absence of parallel pressure, the longitudinal CGL equation \eqref{Macmahoneq} may require further studies when the frozen-in condition for the magnetic field fails to hold and no equation of state is invoked. Hamiltonian structures are now attracting increasing attention in plasma theory due to its benefits especially in particle-in-cell simulations \cite{Krauss,Hong,Eero}. The Hamiltonian structure of certain classes of gyrokinetic moment models was recently studied in \cite{Tassi1,Tassi2}.

Another question that was left open in the present paper is the possibility of quasi-netural drift-ordered models obtained by neglecting the electric energy in the variational principle \cite{CaTr}. In the case of drift-fluid models, this direction was pursued in \cite{Brizard1,Brizard2}, where polarization charges play a crucial role. It would be interesting to know if the introduction of the polarization drift may be avoided in quasi-neutral drift-fluid models or it is instead a necessary ingredient for the consistency of the model.

\paragraph{Acknowledgements.}
I am indebted to Chris Albert, Alain Brizard, Darryl  Holm, Nikita Nikulsin, Stefan Possanner, Bruce  Scott, and Xin Wang for several stimulating discussions on these and related topics. Also, I wish to express my gratitude to Eric Sonnendr\"ucker for his kind hospitality during my stay at IPP. I acknowledge support from the Alexander von Humboldt Foundation (Humboldt Research Fellowship for Experienced Researchers) as well as from the German Federal Ministry for Education  and  Research.

\end{document}